Music Technology Group

Universitat Pompeu Fabra

# When Pamplona sounds different: the soundscape transformation of San Fermín through intelligent acoustic sensors and a sound repository

Amaia Sagasti, Frederic Font

*Versión en Castellano
en Appendix I*

# Acknowledgments

I would like to express my deepest gratitude to Frederic Font, co-author of this paper, for his guidance and support throughout the technical development and creation of the sensors. My thanks also extend to Phonos and the Music Technology Group at Universitat Pompeu Fabra for supporting this project and providing the space and resources that made it possible. I am additionally grateful to KeAcoustics for their valuable technical assistance with the sensors' hardware.

Special thanks are due to the Smart Iruña Lab program of the Pamplona City Council and to all the people who accompanied SENS during this period. Their collaboration was essential not only for the deployment of the sensor network but also for making possible some of the recordings included in the Freesound sound collection.

I would also like to warmly thank my friends and family, whose continuous support encourages every project I undertake. Some of them directly participated in recordings for the sound repository, and I am truly thankful for their help.

Finally, I extend my appreciation to the readers of this paper, whether from Pamplona or beyond; and to those who explore the sound repository, I hope it offers a glimpse into a less-known side of San Fermín.

# Abstract

This study presents a use-case of a network of low-cost acoustic smart sensors deployed in the city of Pamplona to analyse changes in the urban soundscape during the San Fermín Festival. The sensors were installed in different areas of the city before, during, and after the event, capturing continuous acoustic data. Our analysis reveals a significant transformation in the city's sonic environment during the festive period: overall sound pressure levels increase significantly, soundscape patterns change, and the acoustic landscape becomes dominated by sounds associated with human activity. These findings highlight the potential of distributed smart acoustic monitoring systems to characterize the temporal dynamics of urban soundscapes and underscore how the large-scale event of San Fermín Festival drastically reshapes the overall acoustic dynamics of the city of Pamplona. Additionally, to complement the objective measurements, a curated collection of real San Fermín sound recordings has been created and made publicly available, preserving the festival's unique sonic heritage.

# 1. Introduction

Urban soundscapes are dynamic and closely tied to patterns of human activity, mobility, and social interactions. Festivals and large public events represent moments of drastic transformation in these environments, reshaping city rhythms and spatial dynamics. Understanding these changes is crucial for urban planning, noise mitigation, and improving the quality of life in public spaces.

This study focuses on the city of Pamplona/Iruña during the San Fermín Festival, an internationally recognized event that attracts more than a million visitors each year. During nine days, the city undergoes a radical shift in function and activity: streets that usually hold motor traffic become pedestrian areas, while squares and public spaces host dense gatherings, music performances, and celebratory events. These changes produce a distinct acoustic signature compared to the normal conditions of the city.

To explore these transformations, a network of intelligent acoustic sensors was deployed as part of Smart Iruña Lab, a smart city program. These sensors continuously measured sound levels and other acoustic indicators before, during and after the festival. Our analysis addresses how the overall soundscapes in the city change during the San Fermín Festival compared to the non-festival period.

Beyond objective indicators, understanding an event of this magnitude also requires capturing its authentic sonic identity. For this reason, a public repository of characteristic San Fermín sounds has been created on Freesound, providing an accessible record of the festival's intangible auditory heritage. This complementary resource offers a more complete appreciation of the city's transformed soundscape. Throughout the text, links to selected recordings from the repository are provided whenever a described event corresponds to an actual captured sound[1]. Additionally, a dedicated section at the end of this document offers direct access to the full collection.

The measured results show that the San Fermín Festival significantly modifies Pamplona's acoustic environment. Sound pressure levels increase markedly, reflecting increased human presence and social interaction. At the same time, traffic-related sounds diminish due to mobility restrictions and street closures, giving rise to a soundscape dominated by human voices, music, and festive activities. These findings illustrate how distributed sensing technologies can capture the complexity of urban sonic dynamics and support evidence-based approaches to city management.

---

[1] If a hyperlink is not clickable, the corresponding Freesound sound ID (*#ID*) can be accessed manually using the URL format *https://freesound.org/s/<ID>*

## 1.1. Pamplona and San Fermín

Pamplona/Iruña, the capital of Navarre, has about 210,000 inhabitants within the city limits, and over 360,000 including its metropolitan surroundings. Framed by mountains, Pamplona is a medium-sized city — small enough to remain accessible, but large enough to offer cultural variety and commercial vitality. It maintains a modern center with broad avenues and gardens alongside a historic old town walled since medieval times. Family-friendly spaces, sports opportunities, nightlife, and cultural offerings coexist in the daily life of Pamplona.

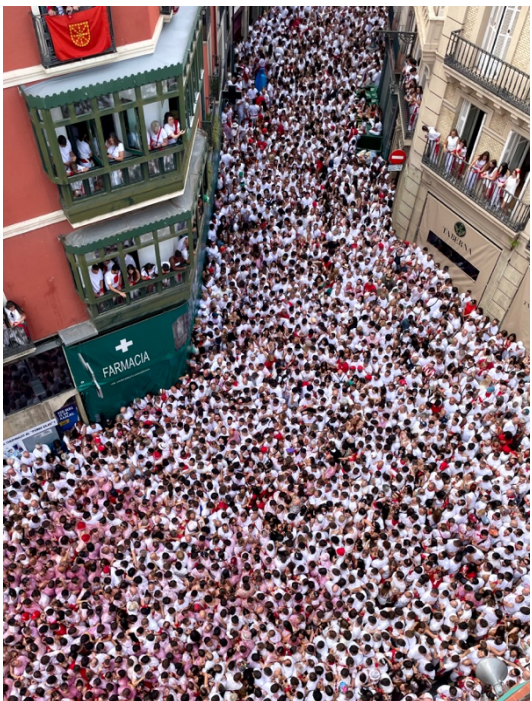

*Figure 1 - Calle Nueva, July 6ᵗʰ, 2025, at noon. Txupinazo event.*

San Fermín is an annual festival held in Pamplona from July 6th to 14th that hosts more than one million visitors each year. Its origins trace back to medieval times, when the relic of Saint Fermín was brought from Amiens to Pamplona in 1186. The *Txupinazo* event (🔊 *#816083*), a firework launched from the balcony of the City Hall, officially begins the festivities at 12:00 on July 6th. Each morning from July 7th to 14th at 8:00, six bulls run through the narrow and crowded Pamplona's streets (🔊 *#816156*), a tradition immortalized by Ernest Hemingway in 'The Sun Also Rises', which brought global fame to the festival.

While bull-related events often steal the outsiders attention, it is music and the vibrant street life that create the heartbeat of San Fermín: *peñas*[2] and their *txarangas*[3] assemble in the streets playing music and energizing the crowd all day long, grand fireworks displays happen every night as part of an international competition and parades of *Gigantes y Cabezudos*[4] are popular within families. Additionally, the festival is renowned for its nightlife — not centered on nightclubs, but on the street gatherings and bars that keep the city awake until dawn. San Fermín unites generations, welcomes visitors from around the world, and weaves music(🔊 *#816150*), ritual(🔊 *#816167*), sport(🔊 *#816078*), and spectacle(🔊 *#816080*) into one resonant sonic environment that transforms the city completely.

## 1.2. The Sensor Device

To monitor the acoustic dynamics of Pamplona before, during and after San Fermín, we used SENS (Smart Environmental Noise System), a low-cost, real-time acoustic monitoring solution designed for urban environments. The entire system was developed in-house, with model training, software engineering, and hardware design and deployment carried out collaboratively by the Music Technology Group[5] and KeAcoustics[6].

SENS continuously captures sound and processes it locally using lightweight artificial intelligence models. Since all processing occurs directly on the device, no audio recordings are permanently stored or transferred, safeguarding privacy. The device not only computes key acoustic parameters, such as sound pressure level (SPL), but also the perceptual and categorical dimensions of the acoustic environment. The ISO-12913[7] standard for the assessment of acoustic environments proposes two key parameters to describe the perceptual quality of soundscapes: *pleasantness*—which reflects how nice or friendly the acoustic scene is—and *eventfulness*—which indicates the degree of activity, with eventful soundscapes characterized by frequent and sudden sounds, and uneventful ones being more constant or monotonous. The SENS sensors are trained to estimate these perceptual sound quality indicators. In addition, they continuously identify the sound sources present in the environment. As SENS is optimized for urban acoustic monitoring, it recognizes typical urban sounds including vehicles, human activity and music. The registered results are transmitted through the wireless network to a remote server for further analysis and clear visualization of the data. Learn more about SENS technology in its dedicated papers[8,9].

---

[2] *Peña*: Social group with its own headquarters and identity (banners, emblems, colours and traditional smocks) that organises activities throughout the year, not only during San Fermín. (🔊 *#816166*)
[3] *Txaranga*: A street marching band with wind and percussion instruments. (🔊 *#816168*)
[4] *Gigantes y cabezudos*: The direct translation is "Giants and big-headed figures". They parade through the streets of Pamplona dancing and playfully chasing and hitting children with foam rods. (🔊 *#816151*)
[5] Music Technology Group, Universitat Pompeu Fabra *https://www.upf.edu/web/mtg*
[6] KeAcoustics. Expert engineering in acoustics, noise and vibrations *https://www.keacoustics.com/*
[7] ISO-12913. *https://www.iso.org/standard/52161.html*

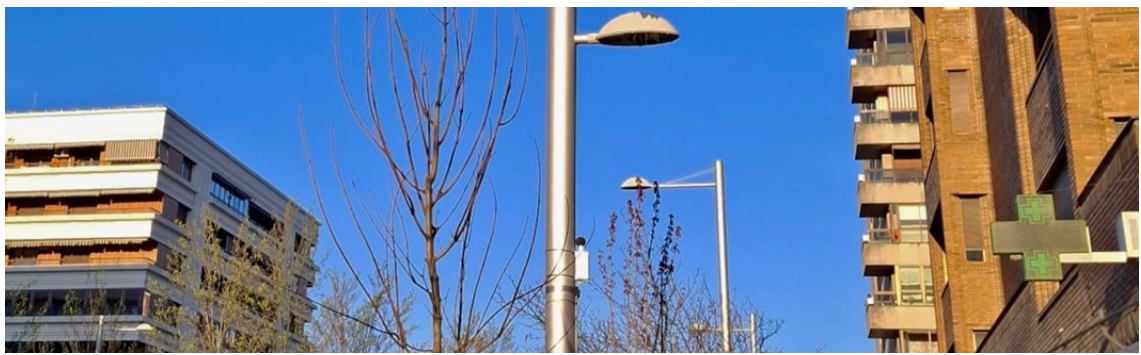

*Figure 2: SENS device deployed in the city of Pamplona/Iruña as part of the Smart Iruña Lab program.*

## 1.3. The Network of Sensors

The deployment of the monitoring network took place in March of 2025 as part of the *Smart Iruña Lab* program, a smart city initiative led by the Pamplona City Council that enables researchers and companies to validate innovative sensor-based solutions in real urban environments. For this study, a network of five SENS devices operated continuously across the city before, during and after the festival, allowing the characterization of soundscape changes associated with this large-scale urban event.

Table 1 summarizes the deployment details. Initially, the five sensors were placed in fixed positions (Spots 1, 3, 4, 6 and 7); however, two of them (Sensors 1 and 3) were relocated on July 3[rd] to areas of greater relevance for the festivities, ensuring comprehensive coverage of the most dynamic soundscapes during San Fermín (Spots 2, 3, 5, 6 and 7). The table also provides an overview of the typical use of these locations under normal conditions.

Spots 2 and 3, both used during and after the festival, do not suffer from any type of restrictions of use due to San Fermín. Similarly, Spots 6 and 7 are kept as pedestrian areas restricted to general traffic, allowing only municipal services, delivery vehicles, or public transport. However, Spot 5, which under normal conditions experiences significant traffic, becomes completely closed to traffic during the festivities, transforming into a pedestrian area. These changes in land use illustrate how urban dynamics are reconfigured during San Fermín, inevitably reshaping the sound environment, a topic examined in detail in the following sections.

*Table 1: Description of the 7 spots where the 5 SENS sensors were deployed during the case study before, during and after the San Fermín festival as part of the Smart Iruña Lab program.*

| Spot | Sensor | Location | Before | During | After | Description |
|------|--------|----------|--------|--------|-------|-------------|
| 1 | 1 | *Calle Irunlarrea* | X | | | Hospital and University area, intermitent traffic |
| 2 | | *Monumento del Encierro* | | X | X | Pedestrian area with commercial activity |
| 3 | 2 | *Avenida Bayona* | X | X | X | Residential area with nightclubs and traffic |
| 4 | 3 | *Rincón de la Aduana* | X | | | Pedestrian area with restricted traffic |
| 5 | | *Labrit con Calle Amaya* | | X | X | Urban area with traffic. |
| 6 | 4 | *Plaza Consistorial* | X | X | X | Pedestrian area with very restricted traffic |
| 7 | 5 | *Paseo Sarasate* | X | X | X | Pedestrian area with restricted traffic |

# 2. Results

From the several months of sensor deployment —some of which were dedicated to refining and calibrating the system — we focus on three specific periods for analysis. The first period, from May 12th to May 18th, represents the city's normal acoustic dynamics during the 'school season', prior to the festival. The second period, July 6th to July 14th, corresponds to the San Fermín Festival, a time of intense urban activity and unique sound patterns. Finally, the third period, from July 15th to July 27th, captures the weeks following the festival, allowing us to assess the transition to the calm city life of the summer.

The following subsections explore two main aspects of the results. First, we perform an in-depth analysis of *Txupinazo* day (July 6th) comparing it to a typical Sunday and examining the SPL at the exact moment of the *Txupinazo* event. Second, we provide an overall comparison of the three selected periods in terms of SPL and other perceptual and categorical parameters, revealing how the festival reshapes the city's soundscape.

## 2.1. Noise Levels on Sunday July 6th

July 6th marks the official beginning of the San Fermín festivities. At exactly 12:00, the traditional *Txupinazo* takes place — a firework that signals the start of the celebration. People typically start the day early, enjoying *el almuercico* (a hearty breakfast of eggs, bacon, fries, and *kalimotxo*[10] as a drink) before heading to the event.

The figures below show the *LAeq* (equivalent continuous A-weighted sound pressure level) values registered on this day, measured at a resolution of one value every 3 seconds, along with the corresponding measurements from two typical Sundays without the festival, May 18th and July 27th.

The *Txupinazo* moment is evident from the sharp increase in sound levels around noon, see Figure 4, with the highest peaks observed at Spot 6, corresponding to *Plaza Consistorial*, the epicenter of this event. This rise in noise levels is noticeable across all locations. Notably, after this peak, the sound levels remain consistently higher than usual

---

[10] *Kalimotxo*: A traditional Basque drink made of red wine and cola.

(compared to the other two Sundays), a pattern that, as we will see in the following section, persists throughout the nine days of the festival.

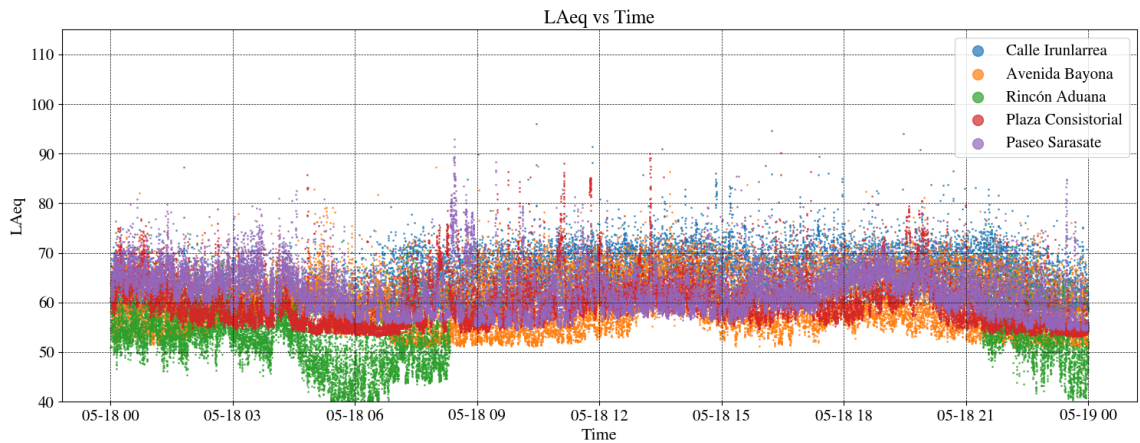

*Figure 3: Raw LAeq captured by the 5 SENS devices on May 18ᵗʰ.*

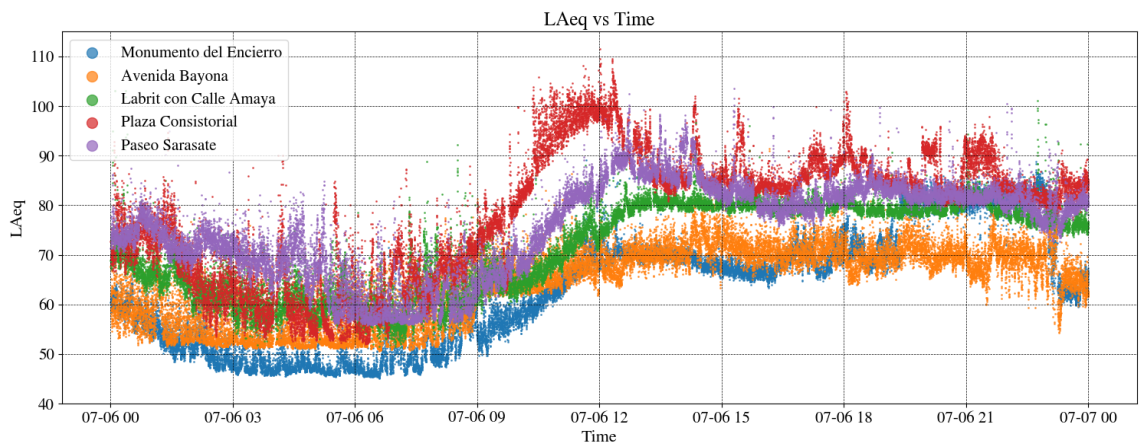

*Figure 4: Raw LAeq captured by the 5 SENS devices on July 6ᵗʰ.*

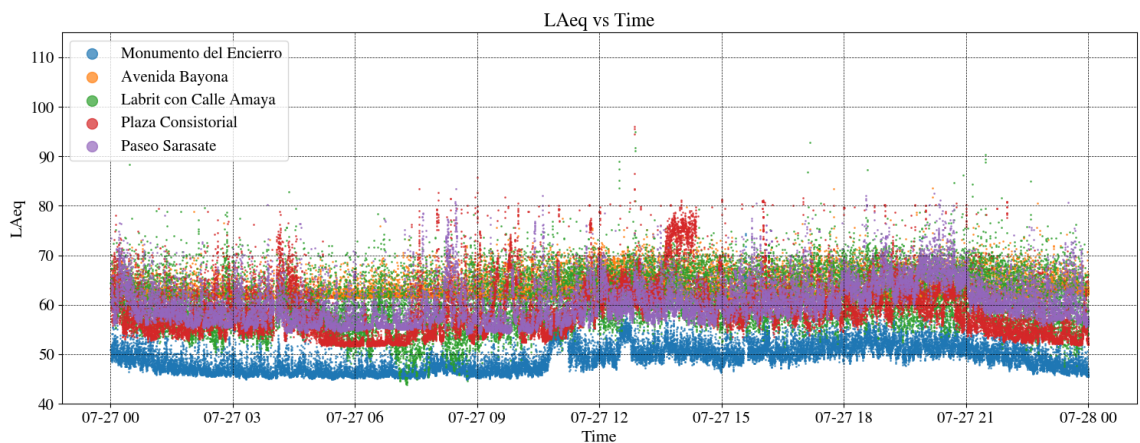

*Figure 5: Raw LAeq captured by the 5 SENS devices on July 27ᵗʰ.*

More informative data can be found in Table 2, which reports the daily *LAeq* value, its difference compared to Sunday July 27th, and the maximum instantaneous *LAeq* measured during *Txupinazo* (considering the period around noon, from 11:58 to 12:05). The maximum value registered is 112 dB at Spot 5, where the *Txupinazo* takes place, close to the upper limits of safe human exposure (120 dB). These results clearly illustrate the significant acoustic impact of the event in terms of noise levels, with an average increase in noise levels of nearly 30% across the five monitored locations.

*Table 2: Summary of SPL values: global values for July 6th and July 27th, and maximum LAeq (3 seconds integration) registered during Txupinazo.*

| Location | LAeq July 6th | LAeq July 27th | Max LAeq *Txupinazo* |
|---|---|---|---|
| *Monumento Encierro* (Spot 2) | 74dB | 51dB | 86dB (at 12:04:34) |
| *Avenida Bayona* (Spot 3) | 69dB | 64dB | 81dB (at 12:02:40) |
| *Labrit, Calle Amaya* (Spot 5) | 78dB | 65dB | 86dB (at 12:01:16) |
| *Plaza Consistorial* (Spot 6) | 90dB | 65dB | 112dB (at 12:01:16) |
| *Paseo Sarasate* (Spot 7) | 82dB | 63dB | 104dB (at 12:01:17) |

## 2.2. Overall Acoustic and Perceptual Characterization of the Festival Period

In this section, we analyze the acoustic environment of the festival period (July 6th to 14th) as a whole, with the aim of identifying pattern changes by comparing it to two non-festival reference periods (May 12th to 18th, and July 15th to 27th). The analysis considers not only the SPL but also the perceptual and categorical dimensions of the acoustic environment, including pleasantness, eventfulness, and the presence of specific sound sources.

To effectively visualize and interpret the data, the raw measurements registered by the devices have been processed according to the criteria summarized in Table 3. Some variables are calculated as averages of the registered values over a defined period, whereas sound sources are expressed as the percentage of time during which a given source was active, or alternatively, as the number of detected events within a period. To determine when a source is considered active, a threshold was manually selected after observation: if the registered values exceed this threshold, the source is active.

To illustrate the daily sound dynamics, circular plots are employed where 360° represent the 24 hours of the day divided into hourly segments (all measurements are computed over one hour periods), and each concentric ring corresponds to a different day. This visualization facilitates the identification of temporal patterns and differences across days and periods. In the following paragraphs, we focus on the most relevant or notable patterns, although the reader is encouraged to explore the full set of graphs for additional insights (see ***Appendix II***).

*Table 3: Brief description on how the monitored parameters results given by the sensor device are interpreted in analysis to extract meaningful data and perform comparisons.*

| Parameter | Sensor output | Interpreted as |
|---|---|---|
| *LAeq* | dBA | Mean values per period |
| *pleasantness* *eventfulness* | Range [-1,1] | Mean values per period, transposed and scaled to the range [0,1] |
| birds, human, vehicles | Range [0,1] | Time percentage above a threshold |
| music | | Count of events |

Starting with the SPL analysis, and as observed in the previous section, the festival period is clearly the noisiest in all monitored locations in comparison to the two non-festival reference periods. Noise levels exceed 80dB for most of the period. Particularly, Sensors 3, 4, and 5, which are located near the festival's epicenter, registered the highest values. Additionally, it is noteworthy that there is little difference in SPL patterns between the pre-festival and post-festival periods. The only exceptions are the measurements from Sensors 1 and 3, but these differences are due to the relocation of these sensors.

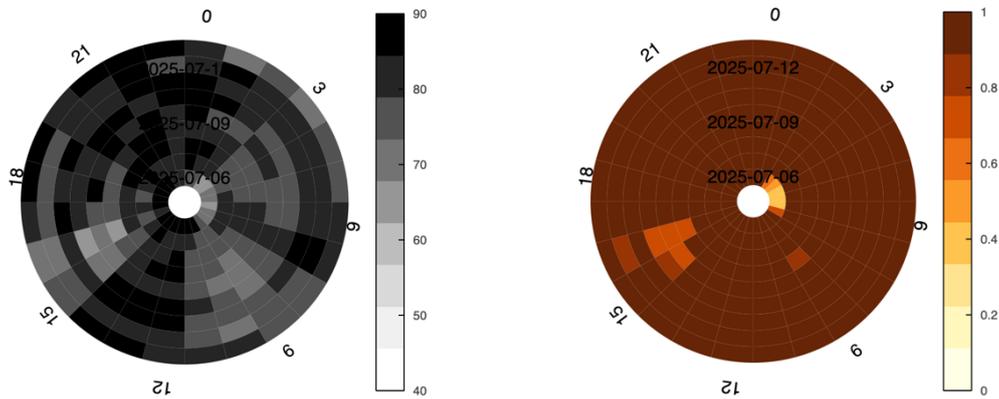

*Figure 6 – LAeq(left) and eventfulness(right) values registered for Sensor 4 (Plaza Consistorial) from July 6th to 14th (at Plaza Consistorial). See more in Figure 15.*

The high SPL values recorded during San Fermín coincide with the similarly elevated levels of eventfulness, see Figure 6. Prominently, eventfulness rises on July 6th around 9:00 and remains high almost continuously until the festival ends. Most locations behave similarly, but we can focus the analysis on Sensor 2 since it reveals some interesting pattern changes, see Figure 7. We observe that during the week of May 12[th] to 18[th], activity levels peak at night (between 02:00 and 05:00) from Thursday to Sunday, reflecting the influence of nearby nightclubs. Additional eventfulness peaks occur on weekends around midday (12:00–14:00) and in the evening (18:00–21:00), consistent with social gatherings. Outside these time periods, eventfulness remains at low-intermediate levels. During San Fermín, however, this pattern changes dramatically. Eventfulness remains high from approximately 09:00 in the morning until 03:00 at night, that is 18 hours of sustained activity, with only a short window of moderate levels during the early morning hours.

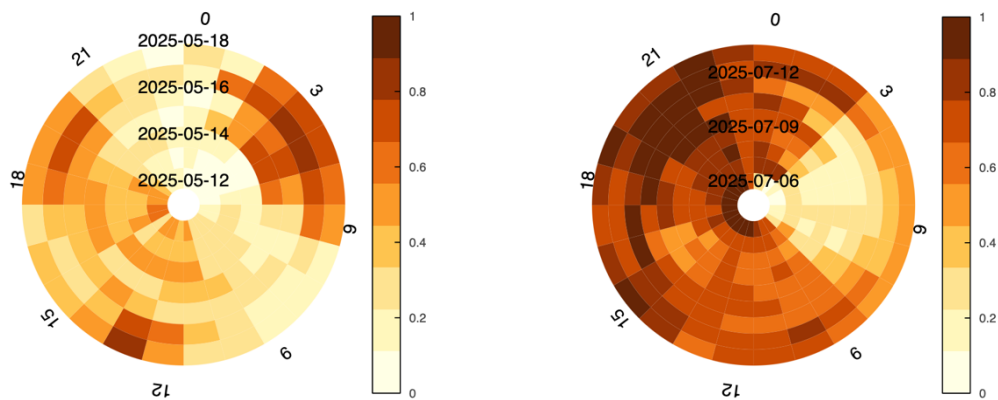

*Figure 7 - Eventfulness values registered for Sensor 2 (Avenida Bayona) from May 12[th] to 18[th] (left) and form July 6[th] to 14[th] (right).*

With respect to pleasantness, the graphs do not reveal changes as dramatic as those observed for eventfulness. Nonetheless, during the non-festival periods, certain hours of the day exhibit noticeably lower pleasantness values. In contrast, during the festival, the pleasantness graph shows a relatively stable pattern of mid-to-high values, visually reflected by a predominantly yellowish to greenish appearance, with very few reddish areas throughout San Fermín, (see graphs d) e) f) from Figure 13, for example). The subjective nature of this parameter makes it difficult to draw strong conclusions about the overall pleasantness of the city's soundscape. However, the apparently more positive perceptual environment during the festival period is likely related to the change in sound sources presence, as discussed in the following paragraphs.

We have observed that SPL and eventfulness levels are substantially higher during San Fermín compared to non-festival periods, but we have not yet explored what drives these increases. Human activity shows a strong correlation with both parameters. Across all locations, human-generated sounds are present for most of the time during the festival — significantly more than in non-festival periods. Sensors 3, 4, and 5, those situated at the core of the festivities, register human activity nearly 100% of the time throughout San Fermín, without interruption (as an example, see Figure 8). This parameter is key and clearly illustrates the acoustic transformation that the city undergoes. It provides an objective indication of how the streets are filled with people, with human presence dominating the soundscape and overshadowing other sound sources.

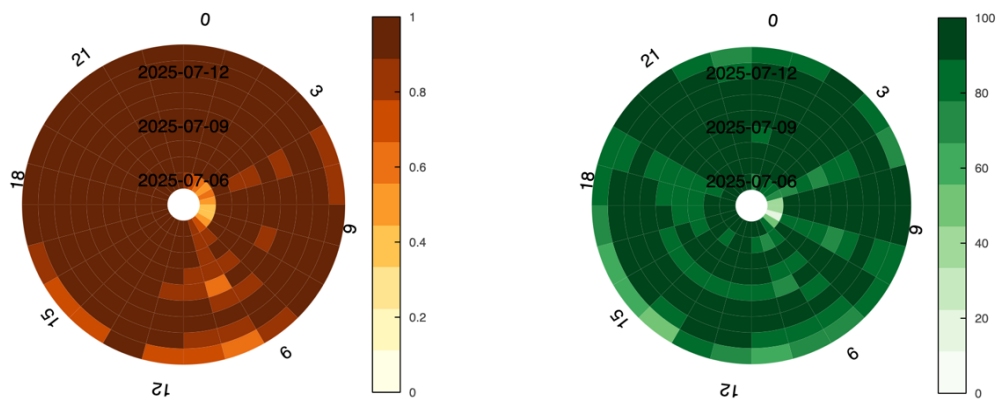

*Figure 8 – Eventfulness(left) and human activity(right) values registered for Sensor 3 (Labrit) from July 6th to 14th. See more in Figure 14.*

The pattern of music presence changes between the festival and non-festival periods, and some interesting trends can be observed at certain locations. For instance, in Figure 8, corresponding to the San Fermín period at Spot 5 (*Cuesta de Labrit*, next to the Bull Ring), clear and repetitive music events are detected at the same times each day: 7:00–8:00, 17:00–18:00, and 21:00–22:00. These patterns correspond to programmed musical events related to the bull activities. The morning music (7:00–8:00) corresponds to the pre-bull run music that is played at the Bull Ring and heard from outside, the early evening music (17:00–18:00) marks the entry of the *peñas* and their *txarangas* to the bullfight event, and the late evening music (21:00–22:00) occurs as they leave the Ring playing music. In this last period, exceptions occur for specific reasons on specific days: no music on the 6th is due to *peñas* not attending the bull-fight, on every 8th, a silent exit from the bullring takes place in memory of a historical event, and this year, no music was played again on the 13ᵗʰ as an act to condemn a sexual assault incident that had occurred the previous day.

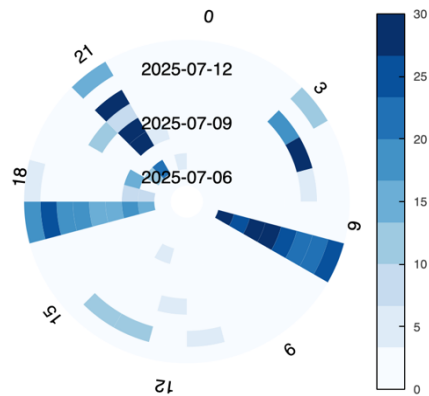

*Figure 9 - Music activity values registered for Sensor 3(Labrit)  from July 6ᵗʰ to 14ᵗʰ. See more in Figure 14.*

Finally, regarding vehicle presence, no notable changes are observed at most locations between the festival and non-festival periods, as traffic restrictions remain largely unchanged. The only exception is Spot 5 (Sensor 3, at *Labrit*), a typically high-traffic area that becomes completely closed to vehicles during the festival. As has been shown in the previous results, this spot lies at the very epicenter of the festivities, where human gatherings and multiple musical events take place. The registered vehicle activity in this spot can be seen in the graphs of Figure 10.

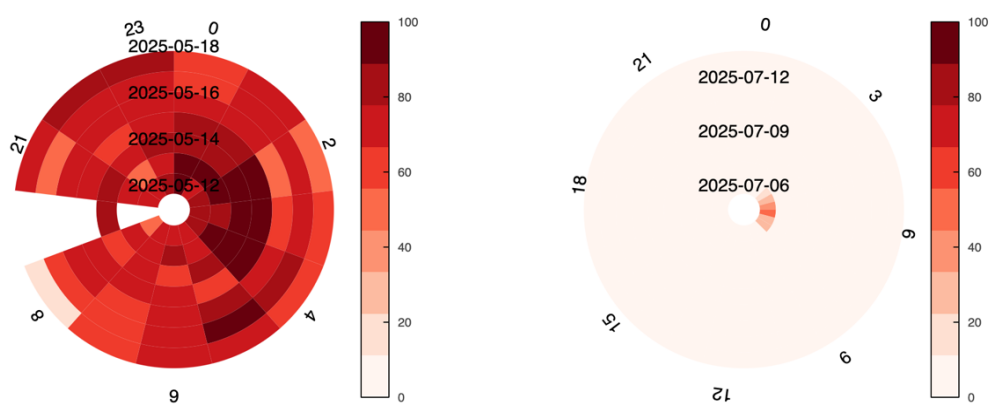

*Figure 10 – Vehicle activity values registered for Sensor 3 (Labrit)  from May 12ᵗʰ to 18ᵗʰ (left) and from July 6ᵗʰ to 14ᵗʰ (right). See more in Figure 14.*

# 3. Discussion

The results presented in this study provide an objective characterization of how San Fermín transforms the acoustic environment of the city of Pamplona/Iruña. By comparing measurements obtained during the nine days of the festival with two non-festival reference periods, clear and consistent patterns emerge across all monitored locations, illustrating the impact that this large event has on the soundscapes of the city.

The most immediate transformation is reflected in the substantial increase in overall sound pressure levels during San Fermín, with notable peaks corresponding to emblematic events such as the *Txupinazo*. These elevated levels coincide with intense social activity and large concentrations of people, conditions that were registered especially at sensors located near the festival's epicentre. The data show that human-generated sounds dominate the acoustic environment throughout the event, often accounting for nearly 100% of active time in central locations. This continuous presence of human activity is one of the most distinctive acoustic signatures of San Fermín.

The analysis of perceptual indicators further reinforces this picture. Eventfulness rises sharply with the start of the festivities and maintains high values for extended periods of the day, reflecting a city in constant motion. In contrast, pleasantness exhibits less variation but remains more positive than during non-festival periods. This suggests that, despite the substantial increase in noise levels, the soundscape is perceived as less negatively valenced, likely due to the dominance of socially meaningful sounds (human presence and music) over mechanical or traffic-related noise.

Overall, these findings highlight how deeply the San Fermín Festival reshapes urban life and how these changes manifest acoustically. A distributed sensing network like SENS enables this transformation to be captured in a comprehensive and quantitative manner. Beyond describing the phenomenon, this analytical framework has practical implications for urban planning and event management. Recognizing temporal patterns in human activity or music events can help authorities anticipate periods of higher crowd density, optimize the deployment of public services, and design targeted interventions for safety and mobility. By enabling evidence-based decision-making, systems like SENS can contribute to designing more resilient, safe, and human-centred urban spaces.

# 4. Beyond the Numbers: Preserving the Sounds of San Fermín

The analysis presented throughout this document is based on objective measurements that capture how the city of Pamplona/Iruña transforms acoustically during San Fermín. These data reveal a soundscape dominated by human presence and music events, elements that shape the city's unique festive atmosphere. However, to truly understand how the city sounds, it is essential to move beyond numerical indicators and incorporate the richness of real-world sonic experiences.

For this reason, supported by the Music Technology Group (Universitat Pompeu Fabra, Barcelona), Phonos (a pioneering centre dedicated to the intersection of sound creation and technological research), and the City Council of Pamplona, I have undertaken an initiative to create an open sound repository of San Fermín within the well-known platform Freesound. Freesound is the world's largest collaborative heterogeneous sound library under Creative Commons (CC) licenses. The platform is built upon the contributions of its community; users freely share audio recordings and choose the licensing terms for their material. This ensures that anyone can listen to and use these sounds, as long as the creators' conditions are respected. In 2025, Freesound celebrates its 20th anniversary with *Intangible Heritage* as its central theme. Intangible heritage encompasses the ephemeral and immaterial elements that shape our world, whether rooted in human culture, natural environments, or the dynamic interplay between them. These fragile and ever-changing sounds are at constant risk of being lost. Through its extensive, community-driven archive, Freesound offers a digital space where such sounds can be preserved, shared, and transformed.

Inspired by this theme, I worked to create a public, accessible archive of many emblematic sounds of the San Fermín Festival. During the 2025 festivities, microphone in hand, I recorded a wide range of events, now available as the *San Fermín[11]* pack on Freesound. The collection includes recordings of songs such as "*Aurora a San Fermín*", the "*Ánimo Pues*" and "*No te Vayas de Navarra*", as well as the sounds of the bull run, the *Txupinazo*, and the nightly fireworks. The goal is to offer an open repository that allows those who have lived the festival to relive it, and those who have not yet had the chance to experience it first-hand to immerse themselves in its atmosphere. Each sound is accompanied by descriptions in English, Spanish, and Basque, along with an image and the recording location.

---

[11] https://freesound.org/people/amaiasagasti/packs/43919/

The next July 6th will bring San Fermín again, with all the emotion and joy that accompanies it every year. Until then, I hope this sonic archive helps us close our eyes and revisit our favourite memories. May you enjoy it as much as I enjoyed creating it.

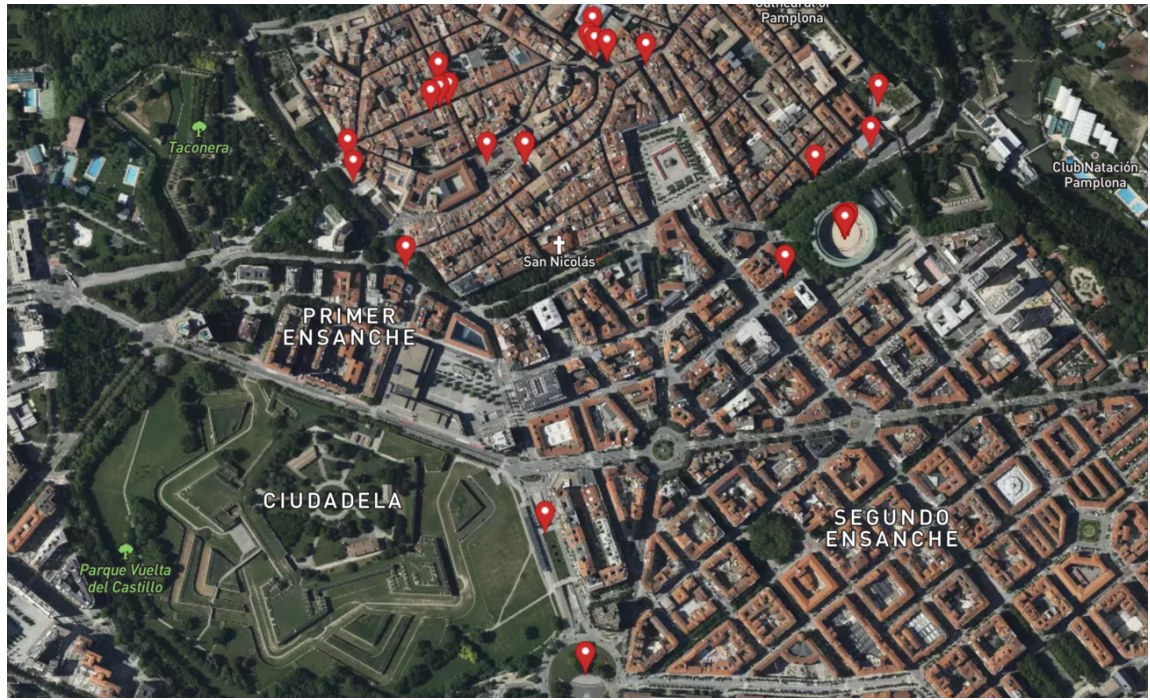

*Figure 11 - Map of sounds for pack 'San Fermín' in Freesound website*

**Appendix I**

Music Technology Group

Universitat Pompeu Fabra

# Cuando Pamplona suena diferente: la transformación del paisaje sonoro en San Fermín a través de sensores acústicos inteligentes y un repositorio sonoro

Amaia Sagasti, Frederic Font

# Agradecimientos

Quiero expresar mi más sincero agradecimiento a Frederic Font, coautor de este trabajo, por su orientación y apoyo durante todo el proceso técnico de desarrollo y creación de los sensores. Mi agradecimiento se extiende también a Phonos y al Music Technology Group de la Universitat Pompeu Fabra por respaldar este proyecto y proporcionar el espacio y los recursos que lo hicieron posible. Asimismo, agradezco a KeAcoustics su valiosa asistencia técnica con el desarrollo del hardware de los sensores.

Quisiera hacer un agradecimiento especial al programa Smart Iruña Lab impulsado por el Ayuntamiento de Pamplona y a todas las personas que acompañaron a SENS durante su participación. Su colaboración fue fundamental no solo para el despliegue de la red de sensores, sino también para hacer posibles algunas de las grabaciones incluidas en la colección de sonidos de Freesound.

Quiero agradecer también de manera especial a mis amigos y familiares, cuyo apoyo constante impulsa e inspira cada proyecto que emprendo. Además, algunos de ellos participaron directamente en las grabaciones del repositorio sonoro, y les estoy profundamente agradecida por su ayuda.

Por último, quiero agradecer a los lectores de este trabajo y a quienes exploren el repositorio sonoro, tanto de Pamplona como de otros lugares; espero que os ofrezca una mirada a un lado menos conocido de San Fermín.

# Resumen

Este estudio presenta un caso de uso de una red de sensores acústicos inteligentes de bajo coste desplegados en la ciudad de Pamplona/Iruña para analizar los cambios en el paisaje sonoro urbano durante las fiestas de San Fermín. Los sensores se instalaron en diferentes puntos de la ciudad antes, durante y después del evento, capturando datos acústicos de manera continua. Nuestro análisis revela una transformación completa en el entorno sonoro de la ciudad durante las fiestas: los niveles de presión sonora aumentan notablemente y los patrones del paisaje sonoro cambian, viéndose dominado por sonidos asociados a la actividad humana. Estos hallazgos destacan el potencial de los sistemas distribuidos de monitorización acústica inteligente para identificar las dinámicas temporales de los paisajes sonoros urbanos, y muestran cómo un evento de gran escala como San Fermín redefine drásticamente la dinámica acústica global de la ciudad de Pamplona. Además, para complementar las mediciones objetivas, se ha creado una colección pública de grabaciones *in situ* de los sonidos de San Fermín, contribuyendo a preservar así el patrimonio sonoro de las fiestas.

# 1. Introducción

Los paisajes sonoros urbanos son dinámicos y están estrechamente ligados a los patrones de actividad humana, movilidad e interacciones sociales. Los grandes eventos públicos representan momentos de transformación de estos entornos, alterando los ritmos y la dinámica espacial de la ciudad. Comprender estos cambios es clave para la planificación urbana, la mitigación del ruido y la mejora de la calidad de vida en los espacios públicos.

Este estudio se centra en la ciudad de Pamplona/Iruña durante las fiestas de San Fermín, un evento de reconocimiento internacional que atrae a más de un millón de visitantes cada año. Durante nueve días, la ciudad experimenta un cambio radical en su función y actividad: calles habitualmente destinadas al tráfico se convierten en zonas peatonales, mientras que plazas y espacios públicos acogen grandes concentraciones de personas, actuaciones musicales y otros eventos festivos. Estas transformaciones generan una huella acústica claramente diferenciada respecto a las condiciones habituales de la ciudad.

Para analizar estos cambios, se desplegó una red de sensores acústicos inteligentes en el marco del programa *Smart Iruña Lab,* una iniciativa de ciudad inteligente del Ayuntamiento de Pamplona. Estos sensores midieron de forma continua los niveles sonoros y otros indicadores acústicos antes, durante y después de las fiestas. El posterior análisis de los datos recogidos muestra cómo el paisaje sonoro global de la ciudad se transforma durante San Fermín en comparación con los periodos no festivos.

Más allá de los indicadores objetivos registrados, comprender un evento de esta magnitud requiere también capturar su identidad sonora auténtica. De esta forma, se ha creado un repositorio público de sonidos característicos de San Fermín en la plataforma Freesound, que ofrece un registro accesible del patrimonio auditivo inmaterial de las fiestas, permitiendo una apreciación más completa del paisaje sonoro de la ciudad. A lo largo del texto se incluyen enlaces a grabaciones seleccionadas del repositorio cuando se menciona un evento concreto[12]. Además, al final de este documento se incluye una sección específica que proporciona acceso directo a la colección completa de grabaciones.

---

[12] Si un enlace no funciona, el identificador de Freesound (*#ID*) del sonido correspondiente puede buscarse manualmente con el formato de URL: https://freesound.org/s/<ID>

Los resultados obtenidos de los sensores muestran que las fiestas de San Fermín modifican de forma significativa el entorno acústico de Pamplona/Iruña. Los niveles de presión sonora aumentan de manera notable, reflejando una mayor presencia humana y una intensa interacción social. Al mismo tiempo, los sonidos asociados al tráfico disminuyen debido a las restricciones de movilidad y al cierre de calles, dando lugar a un paisaje sonoro dominado por voces humanas, música y actividades festivas. Estos hallazgos ilustran cómo redes distribuidas de sensores inteligentes pueden capturar la complejidad de la dinámica sonora urbana para la gestión basada en la evidencia.

## 1.4. Pamplona y San Fermín

Pamplona/Iruña, capital de Navarra, cuenta con unos 210.000 habitantes dentro del término municipal y supera los 360.000 si se incluye su comarca. Rodeada de montañas, Pamplona es una ciudad de tamaño medio: lo suficientemente pequeña como para resultar accesible, pero lo bastante grande como para ofrecer variedad cultural y comercial. Combina un centro moderno, con amplias calles y zonas ajardinadas, con un casco viejo amurallado desde la Edad Media. Espacios familiares, oferta deportiva, vida nocturna y actividades culturales conviven en el día a día de la ciudad.

San Fermín es una fiesta anual que se celebra en Pamplona/Iruña del 6 al 14 de julio y que acoge a más de un millón de visitantes cada año. Sus orígenes se remontan a la Edad Media, cuando en 1186 se trasladó a Pamplona la reliquia de San Fermín desde Amiens, Francia. El evento del *Txupinazo* (🔊 *#816083*), un cohete lanzado desde el balcón del Ayuntamiento, da inicio oficial a las fiestas a las 12:00 del 6 de julio. A partir de ese momento, del 7 al 14 de julio a las 8:00, seis toros recorren las estrechas y abarrotadas calles de Pamplona (🔊 *#816156*), una tradición inmortalizada por Ernest Hemingway en "Fiesta (*The Sun Also Rises*)", dando fama internacional a la celebración.

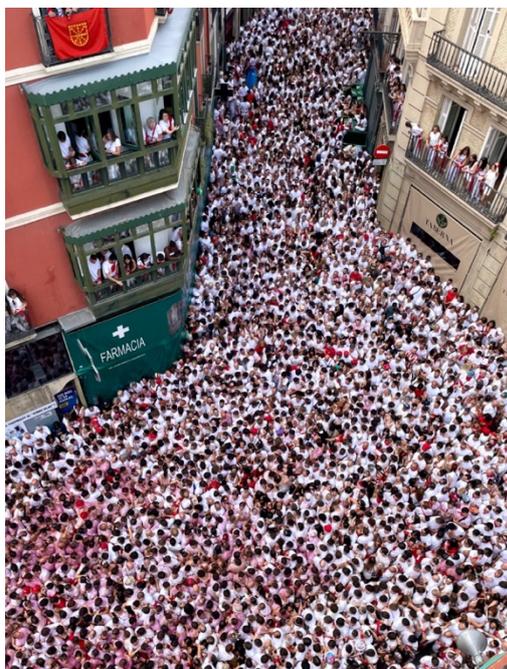

*Figura 1 - Calle Nueva, 6 de Julio de 2025 durante el Txupinazo.*

Aunque los actos taurinos suelen acaparar la atención de quienes visitan la ciudad por primera vez, son la música y la intensa vida en la calle las que constituyen el alma de San Fermín: las peñas[13] y sus *txarangas*[14] se reúnen en las calles animando al público durante todo el día; cada noche tienen lugar grandes espectáculos de fuegos artificiales como parte de un concurso internacional; y los gigantes y cabezudos son especialmente populares entre las familias. Además, la fiesta es conocida por su vida nocturna, que no se basa en discotecas sino en las reuniones en la calle y en los bares que mantienen la ciudad despierta hasta el amanecer. San Fermín une generaciones, acoge a visitantes de todo el mundo y entrelaza música (🔊 *#816150*), religión (🔊 *#816167*), deporte (🔊 *#816078*) y espectáculo (🔊 *#816080*) en un entorno sonoro que transforma por completo la ciudad.

## 1.5. El Sensor

Para monitorizar la dinámica acústica de Pamplona antes, durante y después de San Fermín, se utilizó SENS (*Smart Environmental Noise System*), una solución de monitorización acústica inteligente en tiempo real y de bajo coste diseñada para entornos urbanos. Todo el sistema fue desarrollado internamente: el entrenamiento de los modelos, el desarrollo del software y el diseño e implementación del hardware se llevaron a cabo de forma colaborativa por el Music Technology Group[15] y KeAcoustics[16].

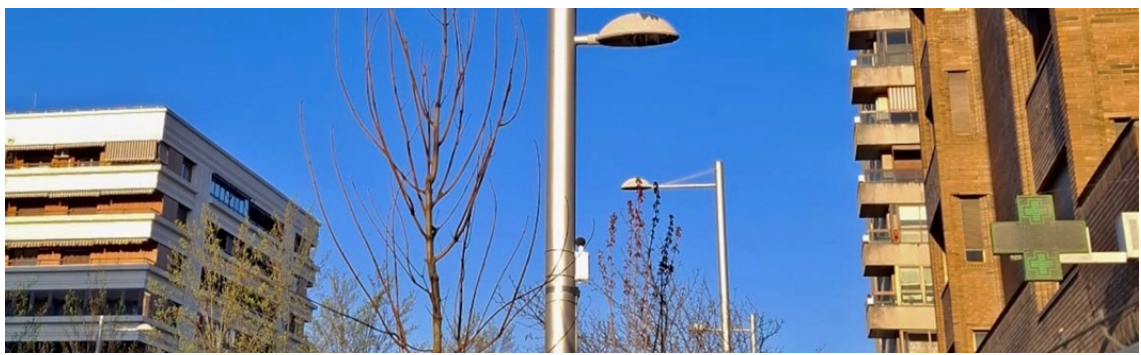

*Figura 2 - Dispositivo SENS en Pamplona/Iruña durante el programa Smart Iruña Lab.*

Cada dispositivo SENS captura el sonido de forma continua y lo procesa localmente mediante modelos ligeros de inteligencia artificial. Dado que todo el procesamiento se realiza directamente en el dispositivo, no se almacenan de manera permanente ni se envían grabaciones de audio, garantizando así la privacidad. El dispositivo no solo calcula el nivel de presión sonora, sino también dimensiones perceptuales y categóricas del entorno acústico. La norma ISO-12913[17] para la evaluación de entornos acústicos propone dos parámetros fundamentales para describir los paisajes sonoros perceptualmente: *pleasantness* (agradabilidad), que refleja cuán agradable o acogedora es la escena acústica, y *eventfulness* (actividad), que indica el grado de dinamismo, considerando más activos aquellos ambientes caracterizados por sonidos frecuentes y repentinos, y poco activos aquellos más monótonos. Los sensores SENS están entrenados para estimar estos indicadores perceptuales. Además, identifican de forma continua las fuentes sonoras del entorno. Al estar optimizado para la monitorización acústica urbana, SENS reconoce sonidos urbanos típicos como el tráfico, la actividad humana y la música. Los resultados registrados se transmiten a través de la red inalámbrica a un servidor remoto para su posterior análisis y visualización de los datos. Para más información sobre la tecnología SENS, pueden consultarse sus publicaciones específicas[18,19].

---

[13] Peña: Asociación social con sede propia e identidad (pancartas, colores y blusas tradicionales) que organiza actividades a lo largo de todo el año, no solo durante las fiestas de San Fermín. (🔊 *#816166*)
[14] *Txaranga*: Banda musical callejera formada por instrumentos de viento y percusión. (🔊 *#816168*)
[15] Music Technology Group, Universitat Pompeu Fabra *https://www.upf.edu/web/mtg*
[16] KeAcoustics. Expert engineering in acoustics, noise and vibrations *https://www.keacoustics.com/*

## 1.6. La Red de Sensores

El despliegue de la red de monitorización tuvo lugar en marzo de 2025 como parte del programa *Smart Iruña Lab*, una iniciativa de ciudad inteligente propulsada por el Ayuntamiento de Pamplona que permite a investigadores y empresas validar soluciones innovadoras en entornos urbanos reales. Para este estudio, una red de cinco dispositivos SENS operó de forma continua en distintos puntos de la ciudad antes, durante y después de las fiestas, lo que permitió caracterizar los cambios en el paisaje sonoro asociados a este evento urbano de gran escala.

La Tabla 1 resume los detalles principales del despliegue. Inicialmente, los cinco sensores se situaron en posiciones fijas (Puntos 1, 3, 4, 6 y 7); sin embargo, dos de ellos (Sensores 1 y 3) fueron reubicados el 3 de julio a zonas de mayor relevancia para las fiestas, garantizando una cobertura adecuada de los paisajes sonoros más dinámicos durante San

---

Fermín (Puntos 2, 3, 5, 6 y 7). La tabla también ofrece una descripción concisa del uso habitual de estas localizaciones en condiciones normales.

*Tabla 1- Descripción de las ubicaciones donde fueron instalados los sensores SENS antes, durante y después de las fiestas de San Fermín, como parte de Smart Iruña Lab.*

| Punto | Sensor | Ubicación | Antes | Durante | Después | Descripción |
|---|---|---|---|---|---|---|
| 1 | 1 | Calle Irunlarrea | X | | | Zona hospitalaria con tráfico intermitente |
| 2 | | Monumento del Encierro | | X | X | Calle peatonal con actividad comercial |
| 3 | 2 | Avenida Bayona | X | X | X | Zona residencial con clubes nocturnos y tráfico |
| 4 | 3 | Rincón de la Aduana | X | | | Zona peatonal con tráfico restringido |
| 5 | | Labrit con Calle Amaya | | X | X | Zona céntrica urbana con abundante tráfico |
| 6 | 4 | Plaza Consistorial | X | X | X | Zona peatonal con tráfico restringido |
| 7 | 5 | Paseo Sarasate | X | X | X | Zona peatonal con tráfico restringido |

Los Puntos 2 y 3, utilizados tanto durante como después de las fiestas, no sufrieron ningún tipo de restricción de uso debido a San Fermín. De manera similar, los Puntos 6 y 7 se mantienen como zonas peatonales con restricciones al tráfico general, permitiéndose únicamente el acceso a servicios municipales, vehículos de reparto o transporte público. En cambio, el Punto 5, que en condiciones normales constituye una zona de alto tránsito de vehículos, queda completamente cerrado al tráfico durante las fiestas, transformándose en un espacio peatonal. Estos cambios en el uso del espacio urbano dan lugar a una reconfiguración en la dinámica de la ciudad durante San Fermín, modificando inevitablemente el entorno sonoro, un aspecto que se analiza en las secciones siguientes.

# 2. Resultados

De los varios meses que duró despliegue de los sensores, algunos de ellos dedicados a la puesta a punto y calibración del sistema, el análisis se centra en tres periodos específicos. El primero, del 12 al 18 de mayo, representa la dinámica acústica habitual de la ciudad durante el "curso escolar", previo al verano y a las fiestas. El segundo periodo, del 6 al 14 de julio, corresponde a San Fermín, etapa de intensa actividad urbana y patrones sonoros especiales. Por último, el tercer periodo, del 15 al 27 de julio, recoge los días posteriores a las fiestas, permitiendo evaluar la transición hacia una rutina más tranquila propia del verano.

Las subsecciones siguientes exploran dos aspectos principales de los resultados. En primer lugar, se realiza un análisis del día del *Txupinazo* (6 de julio), comparándolo con un domingo típico y examinando los niveles de presión sonora (NPS) generales y en el instante del evento del *Txupinazo*. En segundo lugar, se presenta una comparación global de los tres periodos seleccionados en términos de NPS y de otros parámetros perceptivos y categóricos, analizando cómo las fiestas transforman el paisaje sonoro de la ciudad.

## 2.1. Niveles Sonoros el 6 de Julio

El 6 de julio marca el inicio oficial de las fiestas de San Fermín. A las 12:00 tiene lugar el tradicional *Txupinazo*, un cohete que señala el comienzo de la celebración. La jornada suele empezar temprano, con la gente disfrutando del "almuercico" (un desayuno contundente de huevos fritos, panceta, patatas fritas y *kalimotxo*).

Las figuras siguientes muestran los valores de *LAeq* (nivel de presión sonora continuo equivalente ponderado A) registrados durante este día, con una resolución de un valor cada 3 segundos, junto con las mediciones correspondientes a dos domingos sin fiestas, el 18 de mayo y el 27 de julio. El momento del *Txupinazo* se identifica claramente por el brusco aumento de los niveles sonoros en torno al mediodía (véase la Figura 4), con los picos más elevados observados en el Punto 6, correspondiente a la Plaza Consistorial, epicentro del evento. Sin embargo, este incremento de los niveles de ruido es apreciable en todas las localizaciones. Cabe destacar que, tras este pico, los niveles sonoros se mantienen de forma consistente más altos de lo habitual (en comparación con los otros dos domingos), un patrón que, como se verá en la sección siguiente, se prolonga a lo largo de los nueve días de las fiestas.

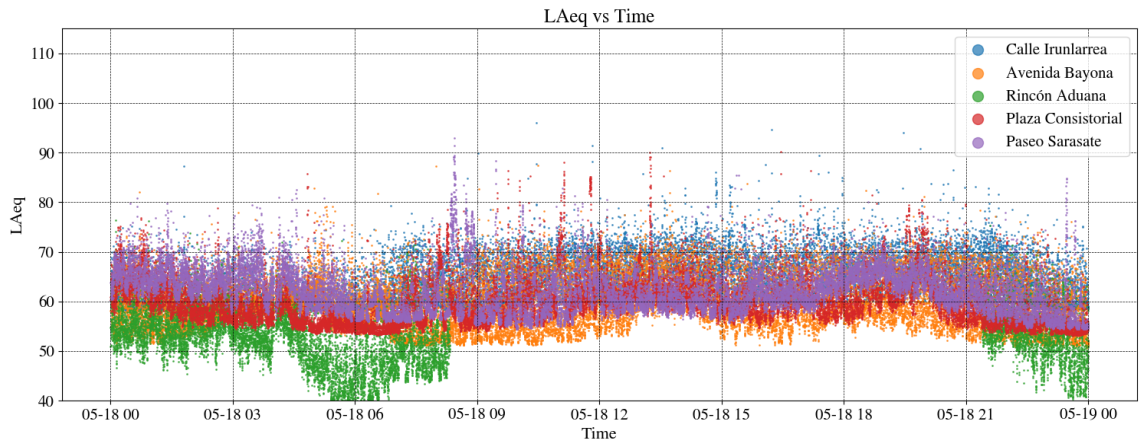

*Figura 3 - Valores de LAeq registrados por los 5 sensores SENS el 18 de mayo*

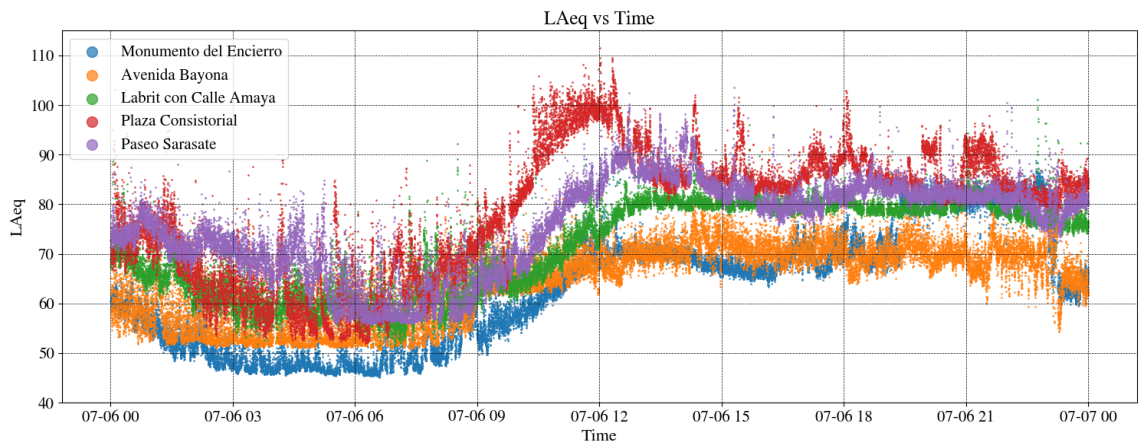

*Figura 4 – Valores de LAeq registrados por los 5 sensores SENS el 6 de julio*

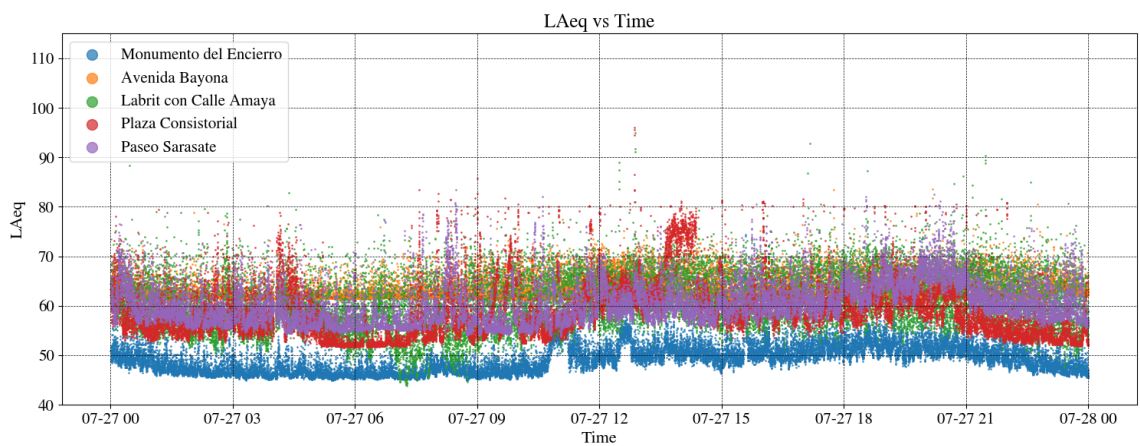

*Figura 5 - Valores de LAeq registrados por los 5 sensores SENS el 27 de julio*

Se puede encontrar información más detallada en la Tabla 2, que recoge el valor diario de *LAeq* de los días 6 y 27 de julio para su comparativa, así como el valor máximo instantáneo de *LAeq* medido durante el *Txupinazo* (considerando el intervalo de 11:58 a 12:05). El valor máximo registrado es de 112 dB en el epicentro del *Txupinazo*, cercano a los límites superiores de exposición segura para el ser humano (120 dB). Estos resultados ilustran el impacto acústico significativo del evento en términos de niveles de ruido, con un aumento medio cercano al 30 % en los cinco puntos monitorizados.

*Tabla 2 - Resumen de NPS: valores globales para el 6 y 27 de julio, y valores máximos (tiempo de intervalo de 3 segundos) registrados en el Txupinazo.*

| Ubicación | *LAeq* 6/julio | *LAeq* 27/julio | Máx. *LAeq Txupinazo* |
|---|---|---|---|
| Monumento Encierro (P.2) | 74dB | 51dB | 86dB (a las12:04:34) |
| Avenida Bayona (P.3) | 69dB | 64dB | 81dB (a las 12:02:40) |
| Labrit, Calle Amaya (P.5) | 78dB | 65dB | 86dB (a las 12:01:16) |
| Plaza Consistorial (P.6) | 90dB | 65dB | 112dB (a las 12:01:16) |
| Paseo Sarasate (P.7) | 82dB | 63dB | 104dB (a las 12:01:17) |

## 2.2. Caracterización Acústica General de San Fermín

En esta sección analizamos el panorama acústico del periodo festivo (del 6 al 14 de julio) en su conjunto, con el objetivo de identificar cambios de patrón comparándolo con dos periodos de referencia sin fiestas (del 12 al 18 de mayo y del 15 al 27 de julio). El análisis considera no solo el nivel de presión sonora (NPS), sino también las dimensiones perceptuales y categóricas del entorno acústico, incluyendo la agradabilidad (*pleasantness*), la actividad (*eventfulness*) y la presencia de fuentes sonoras específicas.

Para poder interpretar los datos de forma práctica, las mediciones en bruto registradas por los dispositivos se han procesado siguiendo los criterios resumidos en la Tabla 3. Algunas variables son calculadas como promedios de los valores registrados durante un periodo definido. Por su lado, las fuentes sonoras identificadas se expresan como el porcentaje de tiempo durante el cual una fuente determinada estuvo activa o, alternativamente, como el número de eventos detectados en un periodo. Para determinar cuándo una fuente se considera activa, se seleccionó manualmente un umbral numérico tras experimentos de observación: si los valores registrados superan dicho umbral, la fuente se considera activa.

Para ilustrar las dinámicas sonoras diarias, se emplean gráficos circulares, en los que los 360° representan las 24 horas del día divididas en segmentos horarios (las mediciones procesadas se calculan sobre periodos de una hora), y cada anillo concéntrico corresponde a un día diferente. Esta visualización facilita la identificación de patrones temporales y diferencias entre días y periodos. En los párrafos siguientes nos centramos en los patrones reconocidos más relevantes o destacables, aunque se anima al lector a explorar el conjunto completo de gráficos para obtener información adicional (ver ***Appendix II***).

*Tabla 3 - Parámetros registrados por SENS y cómo se procesan para su interpretación.*

| Parámetro | Registro bruto | Interpretación |
|---|---|---|
| *LAeq* | dBA | Promedio por periodo |
| *pleasantness* *eventfulness* | Rango [-1,1] | Promedio por periodo, traspuestos y escalados al intervalo [0,1] |
| pájaros, personas, tráfico | Rango [0,1] | Porcentaje de tiempo por encima de un umbral durante un periodo |
| música | | Número de eventos registrados en un periodo |

Comenzando con el análisis del NPS, tal como se ha observado en la sección anterior, el periodo festivo es el más ruidoso en todas las localizaciones en comparación con los dos periodos de referencia sin fiestas. Los niveles superan los 80 dB durante gran parte del periodo. En particular, los Sensores 3, 4 y 5, situados en el epicentro de las fiestas, registran los valores más elevados. Asimismo, es destacable que existen pocas diferencias en los patrones de NPS entre los periodos pre y post-San Fermín. Las únicas excepciones corresponden a los Sensores 1 y 3, las cuales se deben a la reubicación de estos sensores.

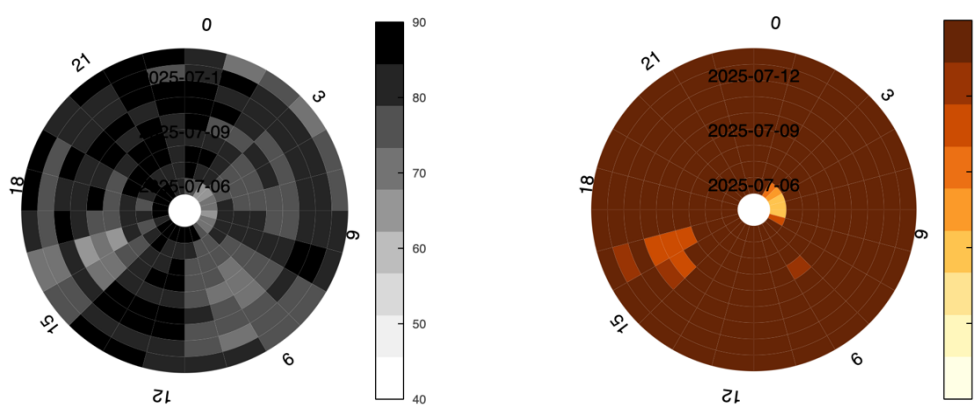

*Figura 6 - Valores de LAeq(izquierda) y de actividad(derecha) registrados por el Sensor 4 del 6 al 14 de julio (Plaza Consistorial). Ver más en Figure 15.*

Los elevados valores de NPS registrados durante San Fermín coinciden con niveles igualmente altos de actividad (*eventfulness*), como se muestra en la Figura 6. De forma destacada, la actividad aumenta el 6 de julio en torno a las 9:00 de la mañana y se mantiene elevada casi de manera continua hasta el final de las fiestas. Aunque la mayoría de las localizaciones presentan comportamientos similares, podemos centrar el análisis en el Sensor 2, ya que revela algunos cambios de patrón especialmente interesantes. Durante la semana del 12 al 18 de mayo, los niveles de actividad alcanzan picos nocturnos (entre las 02:00 y las 05:00) de jueves a domingo, reflejando la influencia de discotecas cercanas. Se observan picos adicionales de actividad los fines de semana alrededor del mediodía (12:00–14:00) y por la tarde (18:00–21:00), relacionados con reuniones sociales. Fuera de estos intervalos, la actividad se mantiene en niveles bajos-intermedios. Sin embargo, durante San Fermín este patrón cambia de forma drástica: la actividad permanece alta aproximadamente desde las 09:00 de la mañana hasta las 03:00 de la noche cada día, es decir, unas 18 horas de actividad sostenida, con solo una breve franja de niveles moderados durante la madrugada (ver Figura 7).

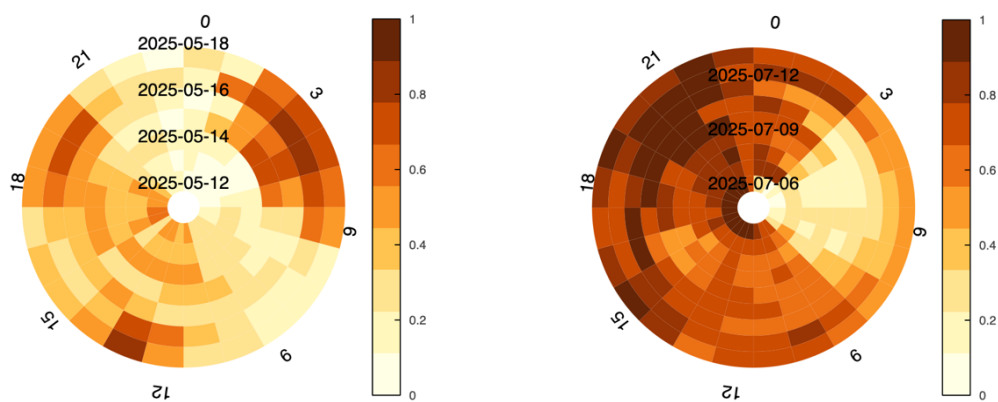

*Figura 7 - Valores de actividad (eventfulness) registrados por el Sensor 2 (Avda. Bayona) del 12 al 18 de mayo (izquierda) y del 6 al 14 de julio (derecha). Ver más en Figure 13.*

En lo que respecta a la agradabilidad (*pleasantness*), los gráficos no muestran cambios tan llamativos como los observados para la actividad. No obstante, durante los periodos sin fiestas, ciertas horas del día presentan valores de agradabilidad notablemente más bajos. En contraste, durante el periodo festivo, el gráfico de agradabilidad muestra un patrón relativamente estable de valores medios a altos, reflejado visualmente por una predominancia de tonos amarillos y verdosos, con muy pocas áreas rojizas en San Fermín (véanse, por ejemplo, los gráficos *d), e)* y *f)* de *Figure 13*). La naturaleza subjetiva de este parámetro dificulta extraer conclusiones contundentes sobre la agradabilidad global del paisaje sonoro de la ciudad. Sin embargo, este entorno perceptual aparentemente más positivo durante las fiestas probablemente esté relacionado con el cambio en la presencia de fuentes sonoras, tal y como se presenta en los párrafos siguientes.

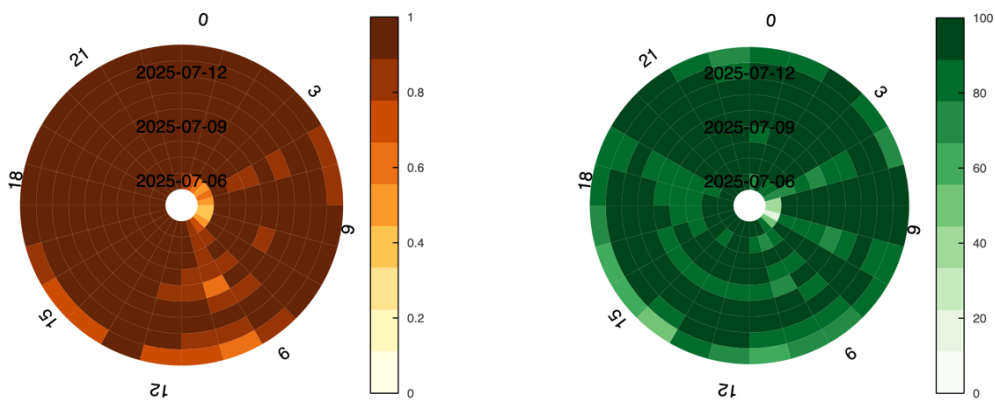

*Figura 8 - Valores de actividad (eventfulness, izquierda) y de actividad humana (derecha) registrados por el Sensor 3 del 6 al 14 de julio (Labrit). Ver más en Figure 14.*

Hemos observado que los niveles de NPS y de actividad son sustancialmente más altos durante San Fermín que en los periodos sin fiestas, pero todavía no hemos explorado qué factores impulsan estos cambios. La actividad humana muestra una fuerte correlación con ambos parámetros. En todas las localizaciones, los sonidos generados por personas están presentes durante la mayor parte del tiempo a lo largo de las fiestas, significativamente más que en los periodos no festivos. Los Sensores 3, 4 y 5, situados en el núcleo de la celebración, registran presencia de actividad humana prácticamente el 100 % del tiempo durante San Fermín, sin interrupciones (véase, como ejemplo, la Figura 8). Este parámetro es clave e ilustra de manera clara la transformación acústica que experimenta la ciudad. Proporciona una indicación objetiva de cómo las calles se llenan de gente, con la presencia humana dominando el paisaje sonoro y eclipsando otras fuentes acústicas.

La presencia de música cambia entre los periodos festivos y no festivos, y se pueden observar tendencias interesantes en determinadas localizaciones. Por ejemplo, en la Figura 9, correspondiente al periodo de San Fermín en la Ubicación 5 (Cuesta de Labrit, junto a la Plaza de Toros), se detectan eventos musicales claros y repetitivos a las mismas horas cada día: de 7:00 a 8:00, de 17:00 a 18:00 y de 21:00 a 22:00. Estos patrones corresponden a eventos musicales programados vinculados a los actos taurinos. La música de la mañana (7:00–8:00) corresponde a la música previa al encierro que suena en la Plaza de Toros y se escucha desde el exterior; la música de primera hora de la tarde (17:00–18:00) indica la entrada de las peñas y sus *txarangas* a la celebración de la corrida; y en la tarde-noche (21:00–22:00) estas salen de la plaza tocando de nuevo. En este último tramo se producen excepciones en días concretos por razones específicas: la ausencia de música el día 6 se debe a que las peñas no participan en el acto taurino; el día 8 tiene lugar una salida silenciosa de la plaza de toros en memoria de un hecho histórico; y este año, además, no se volvió a tocar música el día 13 como acto de condena por una agresión sexual ocurrida el día anterior.

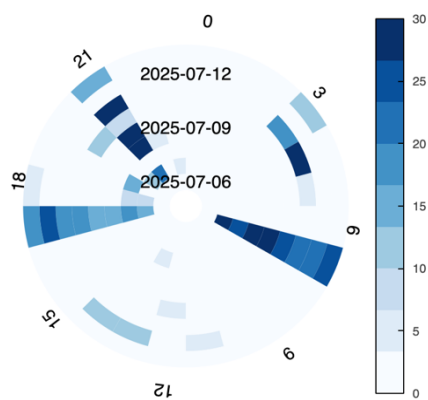

*Figura 9 - Valores de actividad musical registrados por el Sensor 3 del 6 al 14 de julio (Labrit). Ver más en Figure 14.*

Finalmente, en lo que respecta a la presencia de vehículos, no se observan cambios significativos en la mayoría de las localizaciones entre los periodos festivos y no festivos, ya que las restricciones de tráfico se mantienen en gran medida sin variaciones. La única excepción es en el Punto 5 (Sensor 3, en Labrit), una zona que durante las fiestas se cierra completamente a la circulación de vehículos. Como ya se ha mostrado en los resultados anteriores, este punto se sitúa en pleno epicentro de las celebraciones, donde tienen lugar grandes concentraciones de personas y numerosos eventos musicales. Los registros en este emplazamiento pueden observarse en los gráficos de la Figura 10.

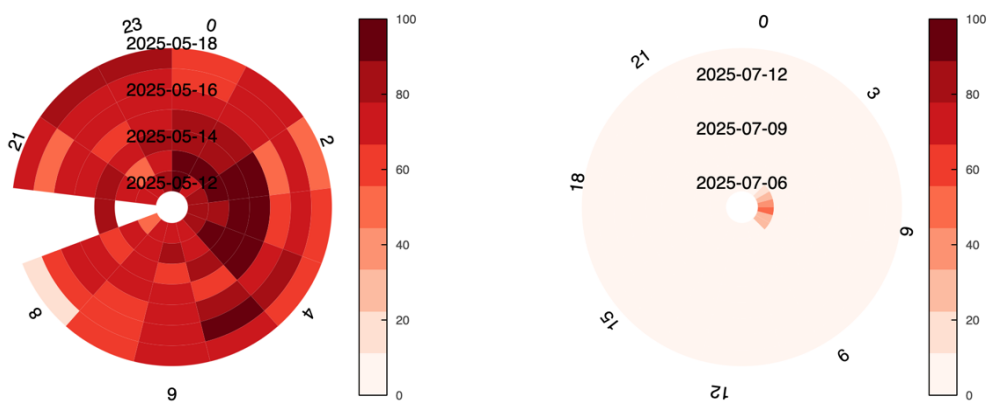

*Figura 10 - Valores de actividad de vehículos registrados para el Sensor 3 (Labrit) del 12 al 18 de mayo (izquierda) y del 6 al 14 de julio (derecha). Ver más en Figure 14.*

# 3. Conclusiones

Los resultados presentados ofrecen una caracterización objetiva de cómo las fiestas de San Fermín transforman el entorno acústico de la ciudad de Pamplona/Iruña. Al comparar las mediciones obtenidas durante los nueve días de las fiestas con dos periodos de referencia sin festividad, emergen patrones en las localizaciones monitorizadas que ilustran el impacto que este gran evento tiene sobre los paisajes sonoros de la ciudad.

La transformación más llamativa se refleja en el aumento sustancial de los niveles globales de presión sonora durante San Fermín, con picos destacados asociados a eventos emblemáticos como el *Txupinazo*. Estos niveles elevados coinciden con una intensa actividad social, condiciones que se registraron especialmente en los sensores situados cerca del epicentro de las fiestas. Los datos muestran que los sonidos generados por personas dominan el entorno acústico a lo largo de todo el evento, llegando a casi un 100% de tiempo de actividad en los núcleos principales. Esta presencia continua de actividad humana constituye una de las firmas acústicas más distintivas de San Fermín.

El análisis de los indicadores perceptuales refuerza aún más este panorama. Los niveles de actividad (*eventfulness*) aumentan de forma abrupta con el inicio de las fiestas y se mantienen en valores elevados, reflejando una ciudad en movimiento constante. En contraste, la agradabilidad (*pleasantness*) presenta variaciones menos extremas, pero se registran valores más positivos a los observados durante los periodos no festivos. Esto sugiere que, a pesar del incremento de los niveles de ruido y de actividad, el paisaje sonoro se percibe más amigable, probablemente debido al predominio de sonidos humanos (presencia de personas y música) frente a ruidos mecánicos o relacionados con el tráfico.

En conjunto, estos resultados ponen de manifiesto hasta qué punto las fiestas de San Fermín reconfiguran la vida urbana y cómo estos cambios se manifiestan acústicamente. Una red de sensores distribuida como SENS permite capturar esta transformación de manera cuantitativa. Más allá de la mera descripción del fenómeno, este análisis puede tener implicaciones prácticas para la planificación urbana y la gestión de eventos. SENS contribuye a la toma de decisiones basada en la evidencia y puede ayudar a las autoridades a anticipar periodos de mayor afluencia de personas, optimizar el despliegue de servicios públicos y diseñar intervenciones específicas en materia de seguridad y movilidad.

# 4. Más Allá de los Números: Preservando los Sonidos de San Fermín

El análisis presentado se basa en mediciones objetivas que capturan cómo la ciudad de Pamplona/Iruña se transforma acústicamente durante las fiestas de San Fermín. Estos datos revelan un paisaje sonoro dominado por la presencia humana y los eventos musicales, elementos que configuran una atmósfera festiva única. Sin embargo, para comprender verdaderamente cómo suena la ciudad, es fundamental ir más allá de los indicadores numéricos e incorporar la riqueza de las experiencias sonoras reales.

Por este motivo, contando con el apoyo del Music Technology Group (Universitat Pompeu Fabra, Barcelona), de Phonos (centro pionero dedicado a la intersección entre la creación sonora y la investigación tecnológica) y del Ayuntamiento de Pamplona, he llevado a cabo la iniciativa de crear un repositorio abierto de sonidos de San Fermín dentro de la conocida plataforma Freesound. Freesound constituye la mayor biblioteca colaborativa de sonidos heterogéneos del mundo bajo licencias *Creative Commons* (CC). La plataforma se sustenta en las aportaciones de su comunidad: los usuarios comparten libremente grabaciones de audio y deciden las condiciones de licencia de su material. De este modo, cualquiera puede escuchar y utilizar estos sonidos, siempre que se respeten las condiciones establecidas por sus autores. En 2025, Freesound celebra su 20º aniversario con el "Patrimonio Inmaterial" (*Intangible Heritage*) como tema central. El patrimonio inmaterial abarca los elementos efímeros e intangibles que conforman nuestro mundo, ya estén arraigados en la cultura humana, en los entornos naturales o en la interacción dinámica entre ambos. Estos sonidos frágiles y cambiantes corren el riesgo de desaparecer. A través de su extenso archivo, Freesound ofrece un espacio digital para la comunidad donde estos sonidos pueden preservarse, compartirse y transformarse.

Inspirada por esta temática, he trabajado en la creación de un archivo público y accesible de muchos de los sonidos más emblemáticos de las fiestas de San Fermín. Durante los Sanfermines de 2025, con micrófono en mano, realicé grabaciones de una variedad de eventos que ahora están disponibles para escuchar en Freesound en el *pack* "San Fermín"[20]. La colección incluye canciones como "Aurora a San Fermín", el "Ánimo Pues" o "No te Vayas de Navarra", así como el sonido del encierro, del *Txupinazo* y de los fuegos artificiales nocturnos. El objetivo es ofrecer un repositorio que permita revivir la fiesta a quienes ya hemos tenido la suerte de vivirla y que, al mismo tiempo, permita a quienes aún no han tenido la oportunidad de experimentarla en primera persona sumergirse en su especial atmósfera. Cada sonido va acompañado de descripciones en inglés, castellano y euskera, así como de una imagen y de la localización de la grabación.

---

[20] https://freesound.org/people/amaiasagasti/packs/43919/

El próximo 6 de julio volverá San Fermín, con toda la emoción y la alegría que lo acompañan cada año. Hasta entonces, espero que este archivo sonoro os ayude a cerrar los ojos y a revivir vuestros recuerdos favoritos. Ojalá lo disfrutéis tanto como yo he disfrutado creándolo.

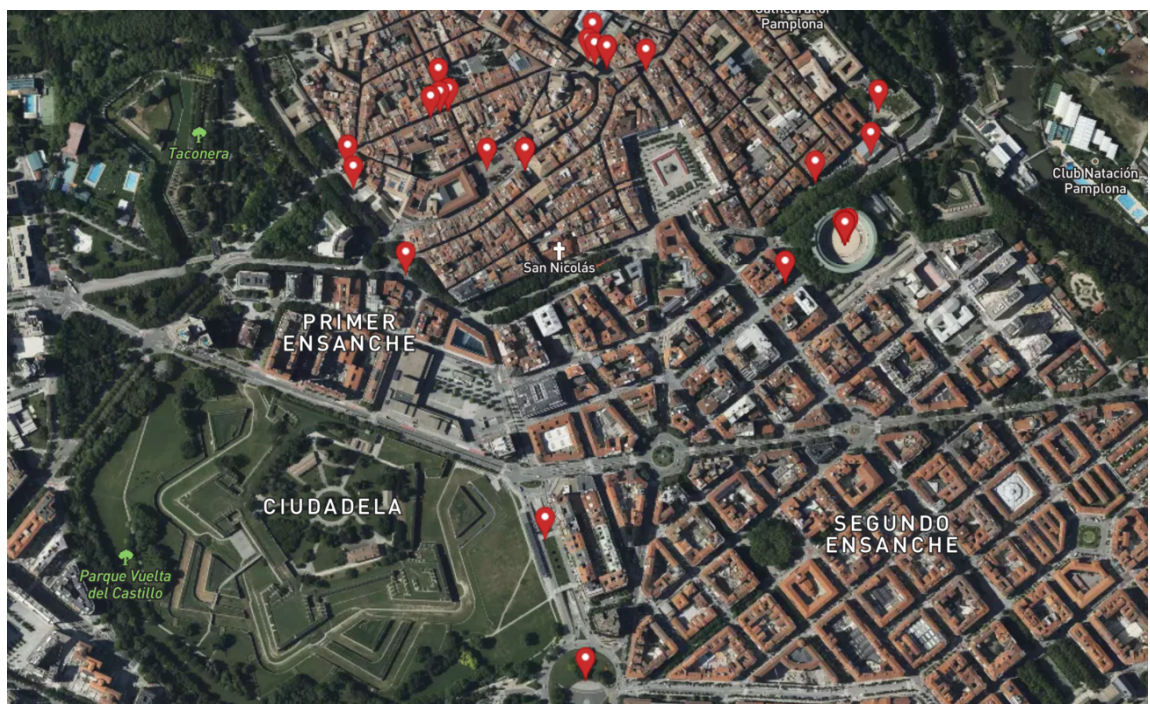

*Figura 11 - Mapa de sonidos del pack San Fermín en la web Freesound.*

# Appendix II

*Figure 12 - Sensor 01 circular plots across time and locations.*

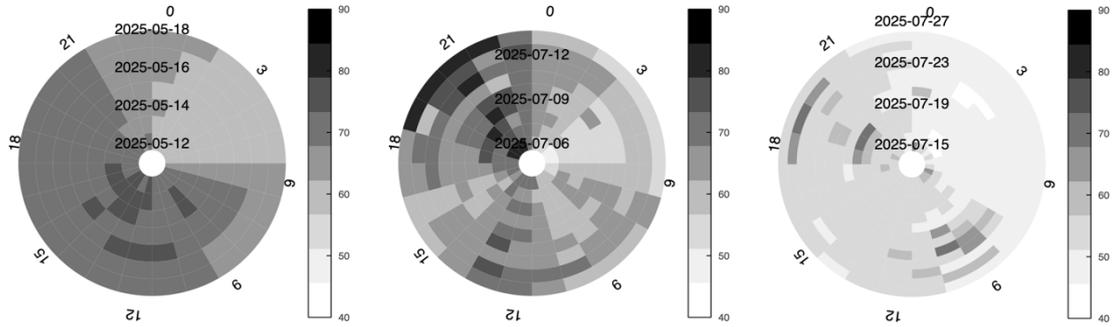

*a) LAeq, May 12-18th*        *b) LAeq, July 6-14th*        *c) LAeq, July 15-27th*

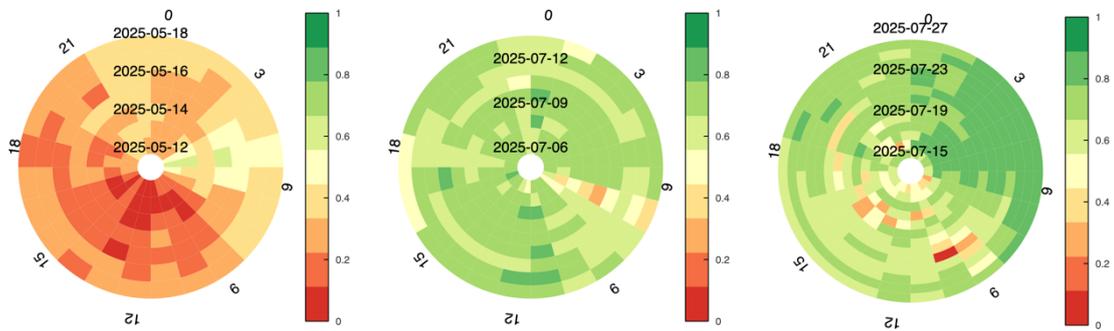

*d) Pleasantness, May 12-18th*   *e) Pleasantness, July 6-14th*   *f) Pleasantness, July 15-27th*

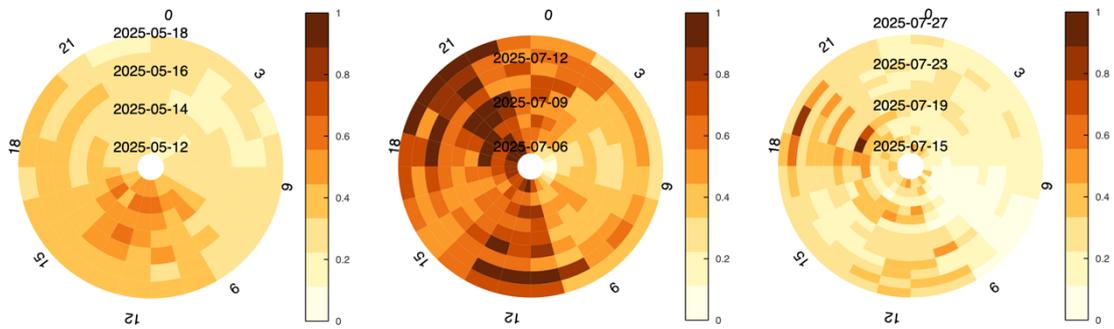

*g) Eventfulness, May 12-18th*   *h) Eventfulness, July 6-14th*   *i) Eventfulness, July 15-27th*

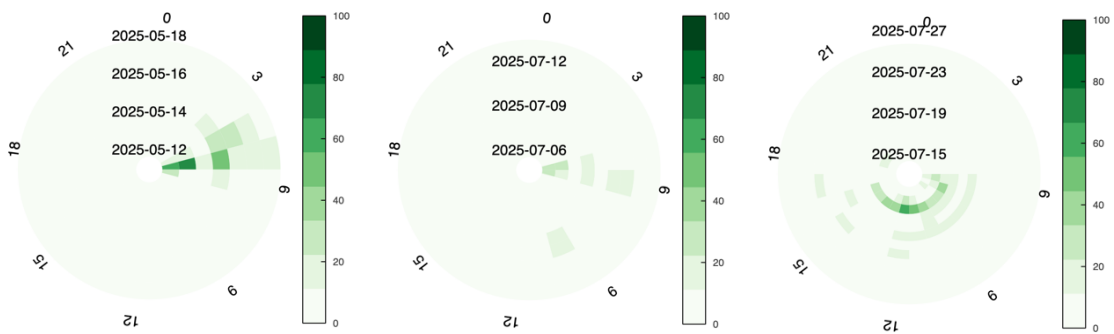

*j) Birds, May 12-18th*        *k) Birds, July 6-14th*        *l) Birds, July 15-27th*

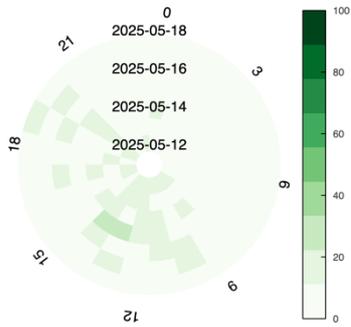
*m) Human, May 12-18th*

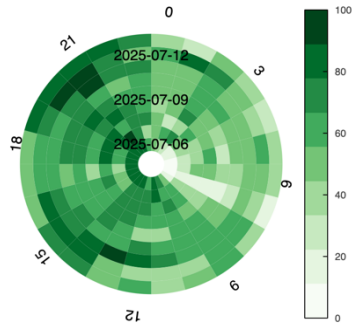
*n) Human, July 6-14th*

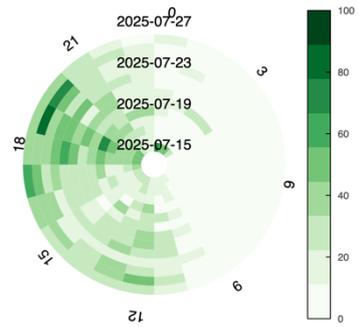
*o) Human, 15-27th*

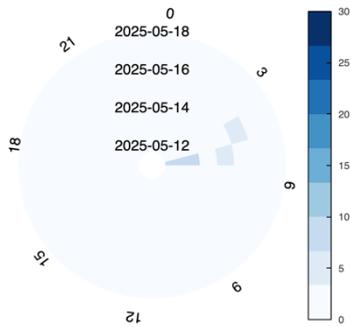
*p) Music, May 12-18th*

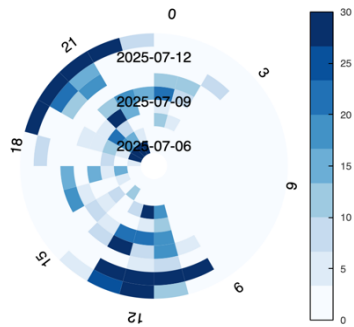
*q) Music, July 6-14th*

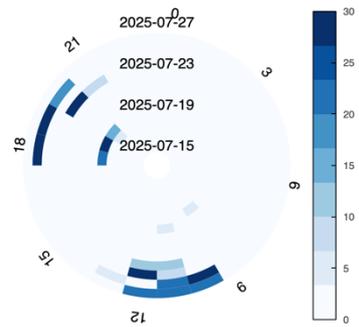
*r) Music, July 15-27th*

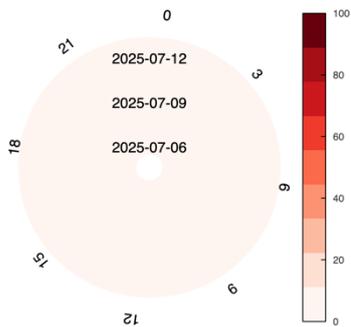
*s) Vehicles, May 12-18th*

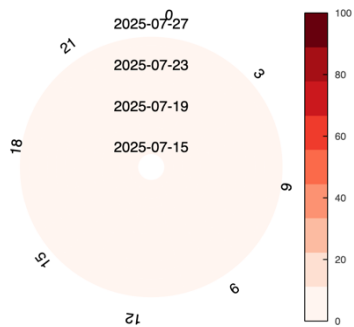
*t) Vehicles, July 6-14th*

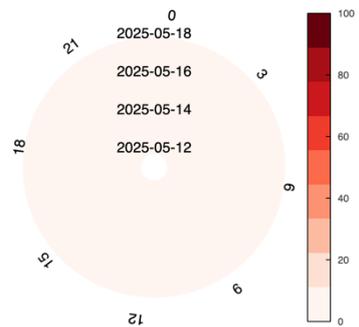
*u) Vehicles, July 15-27th*

Go to Results    Ir a Resultados

*Figure 13 - Sensor 02 circular plots across time and locations.*

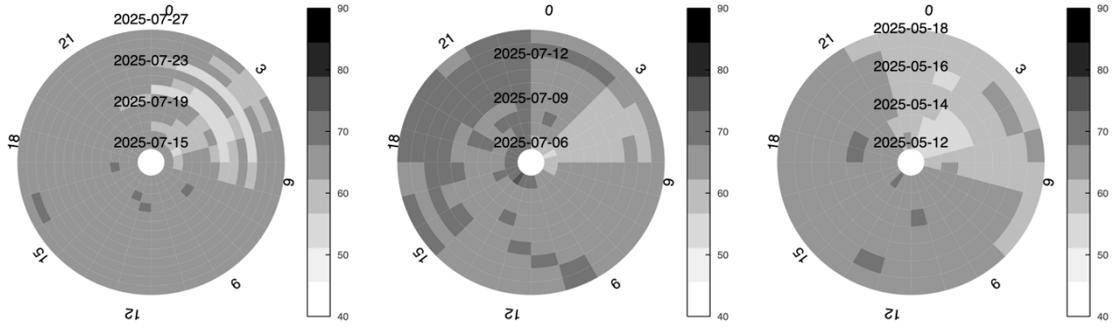

*a) LAeq, May 12-18th*    *b) LAeq, July 6-14th*    *c) LAeq, July 15-27th*

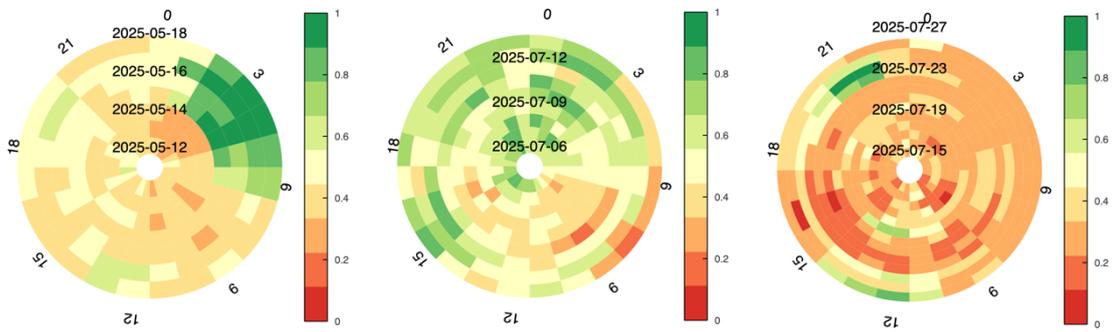

*d) Pleasantness, May 12-18th*    *e) Pleasantness, July 6-14th*    *f) Pleasantness, July 15-27th*

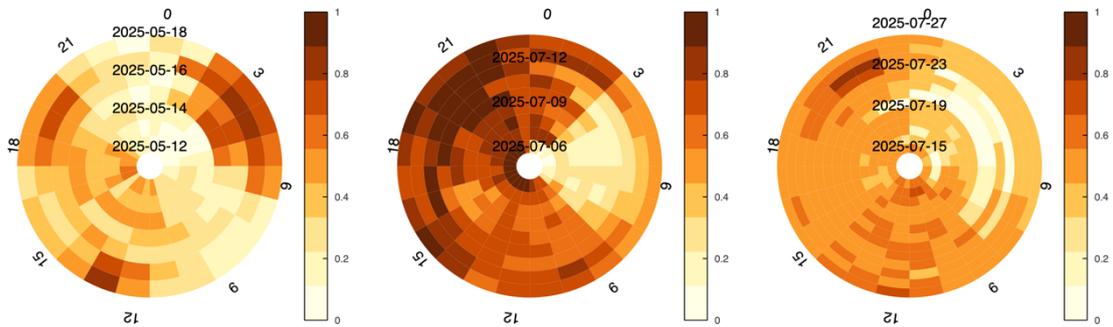

*g) Eventfulness, May 12-18th*    *h) Eventfulness, July 6-14th*    *i) Eventfulness, July 15-27th*

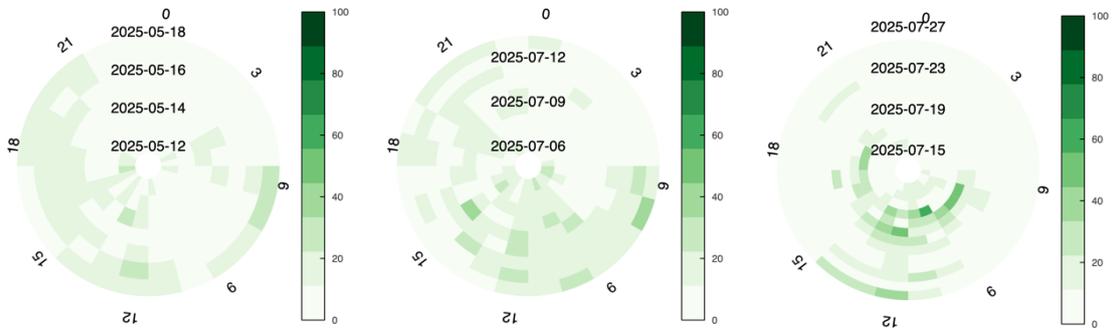

*j) Birds, May 12-18th*    *k) Birds, July 6-14th*    *l) Birds, July 15-27th*

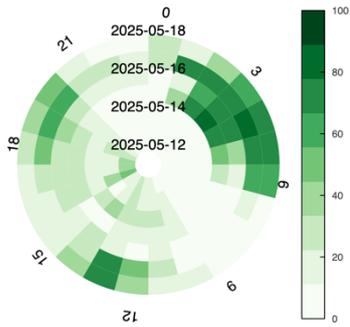

*m) Human, May 12-18th*

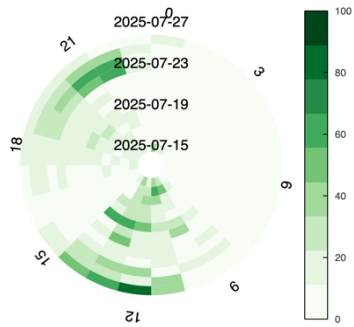

*n) Human, July 6-14th*

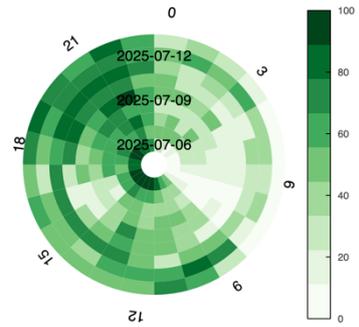

*o) Human, 15-27th*

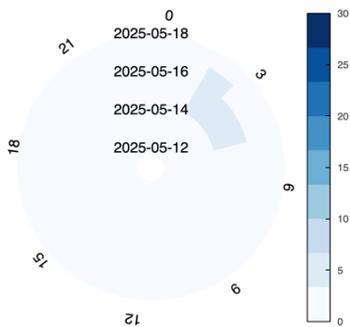

*p) Music, May 12-18th*

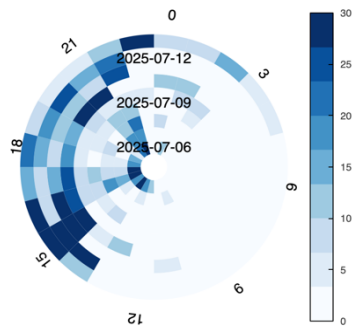

*q) Music, July 6-14th*

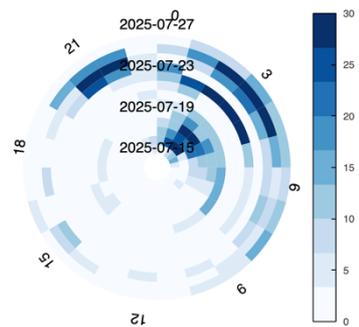

*r) Music, July 15-27th*

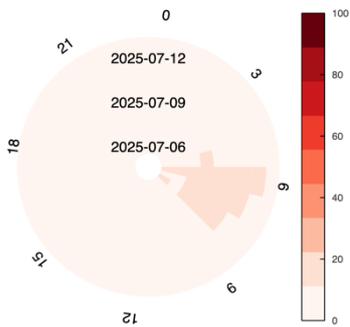

*s) Vehicles, May 12-18th*

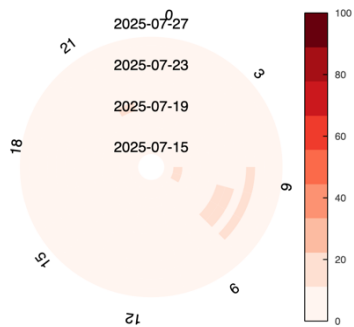

*t) Vehicles, July 6-14th*

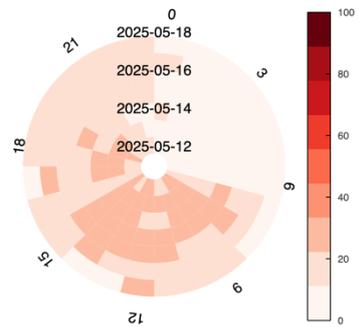

*u) Vehicles, July 15-27th*

Go to Results    Ir a Resultados

*Figure 14 - Sensor 03 circular plots across time and locations.*

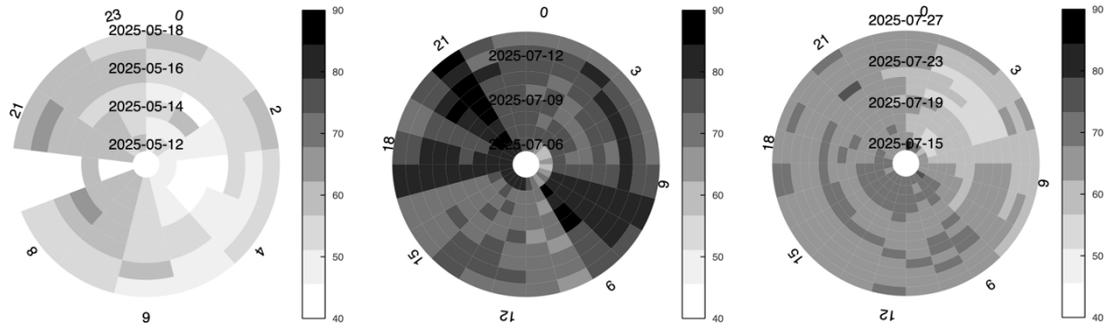

*a) LAeq, May 12-18th*    *b) LAeq, July 6-14th*    *c) LAeq, July 15-27th*

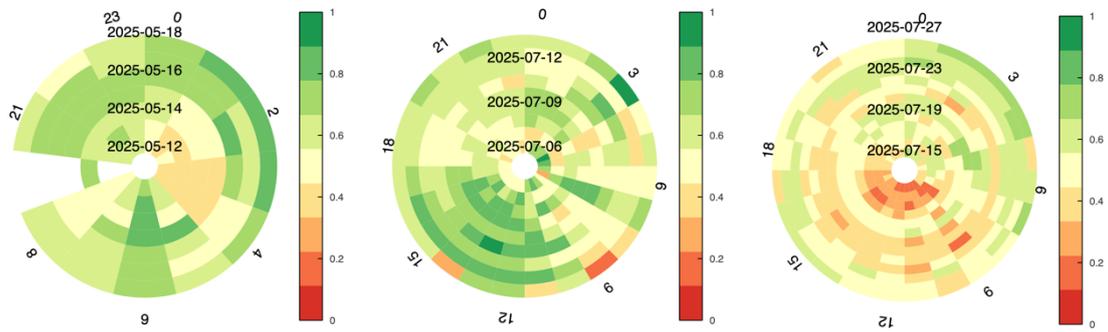

*d) Pleasantness, May 12-18th*    *e) Pleasantness, July 6-14th*    *f) Pleasantness, July 15-27th*

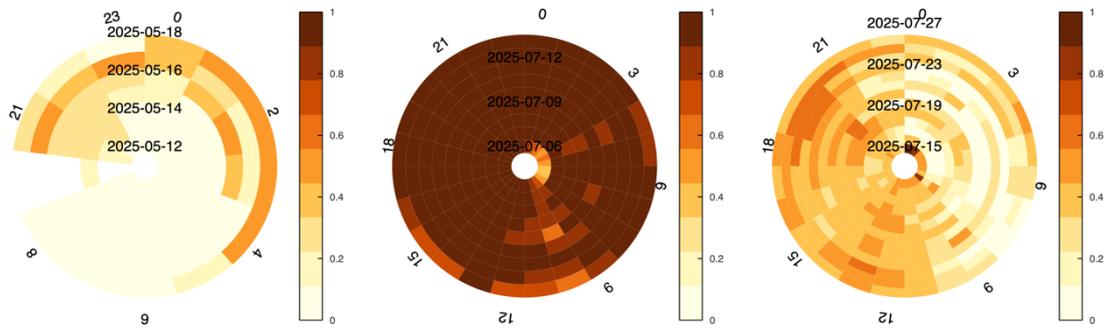

*g) Eventfulness, May 12-18th*    *h) Eventfulness, July 6-14th*    *i) Eventfulness, July 15-27th*

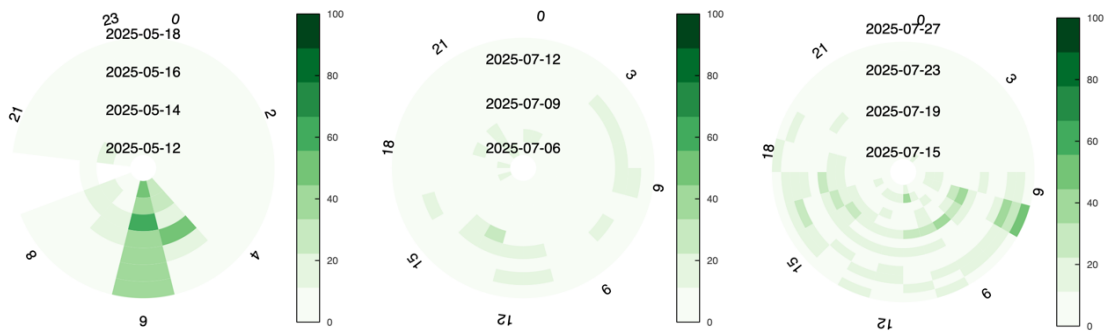

*j) Birds, May 12-18th*    *k) Birds, July 6-14th*    *l) Birds, July 15-27th*

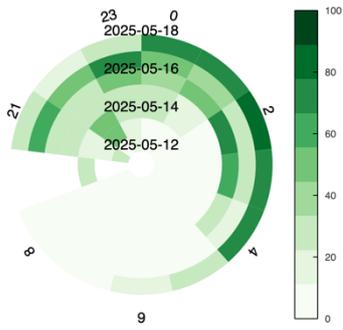

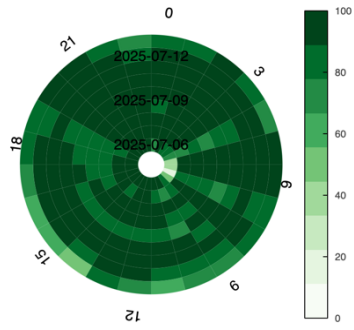

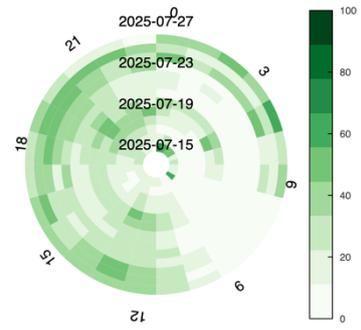

*m) Human, May 12-18th*     *n) Human, July 6-14th*     *o) Human, 15-27th*

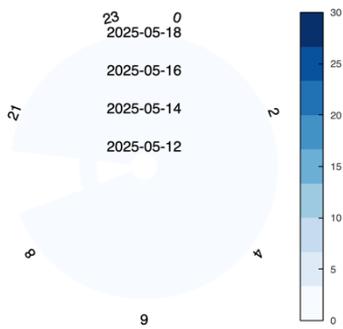

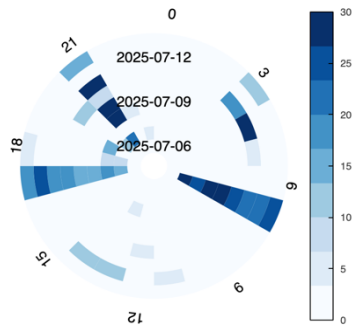

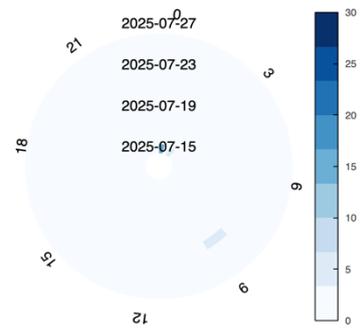

*p) Music, May 12-18th*     *q) Music, July 6-14th*     *r) Music, July 15-27th*

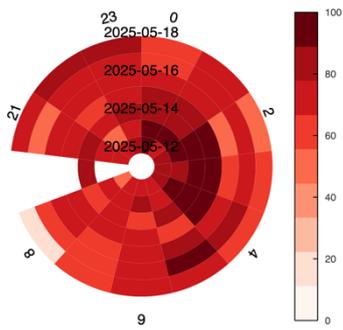

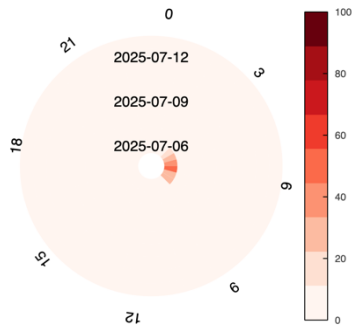

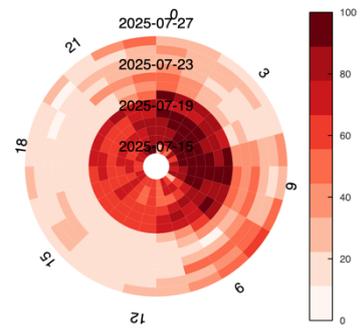

*s) Vehicles, May 12-18th*     *t) Vehicles, July 6-14th*     *u) Vehicles, July 15-27th*

Go to Results     Ir a Resultados

*Figure 15 - Sensor 04 circular plots across time and locations.*

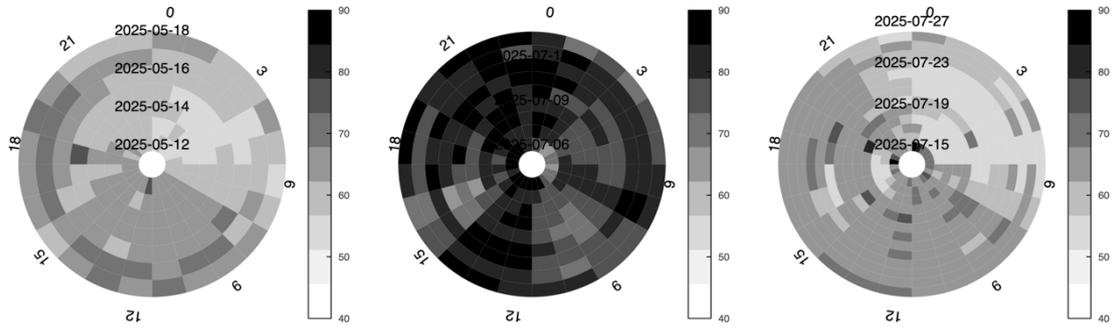

*a) LAeq, May 12-18th*    *b) LAeq, July 6-14th*    *c) LAeq, July 15-27th*

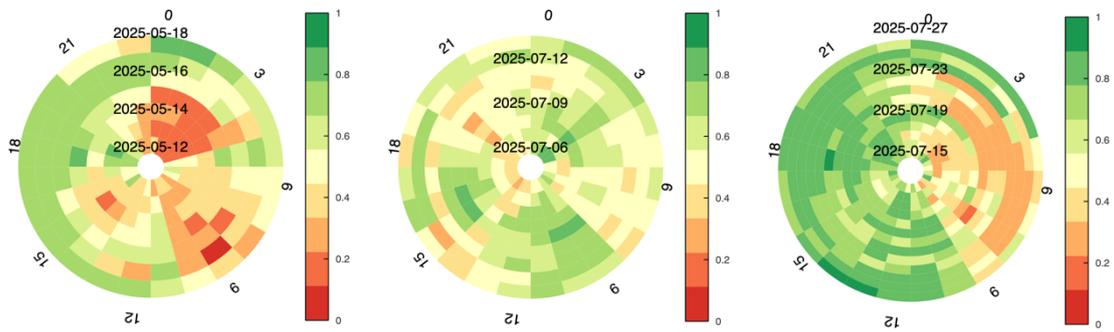

*d) Pleasantness, May 12-18th*    *e) Pleasantness, July 6-14th*    *f) Pleasantness, July 15-27th*

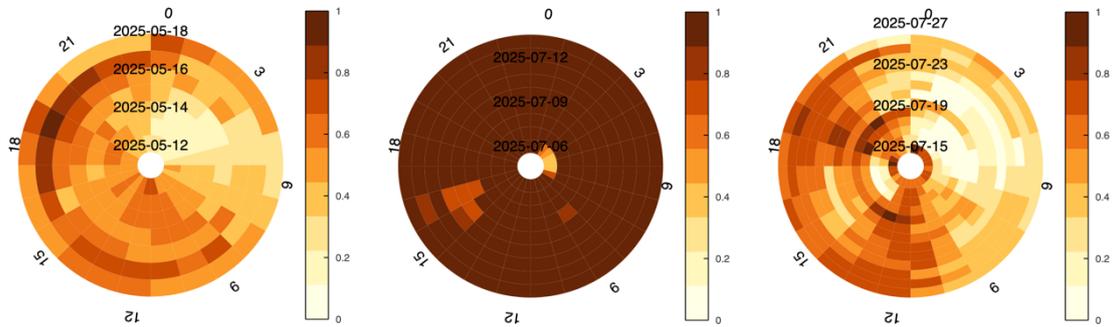

*g) Eventfulness, May 12-18th*    *h) Eventfulness, July 6-14th*    *i) Eventfulness, July 15-27th*

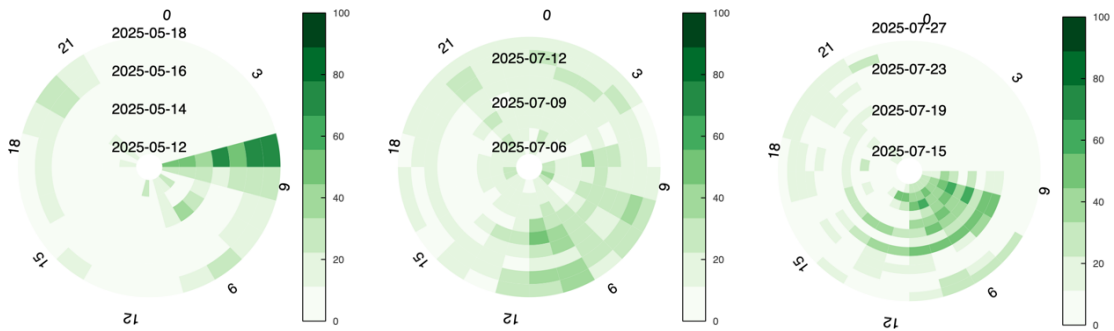

*j) Birds, May 12-18th*    *k) Birds, July 6-14th*    *l) Birds, July 15-27th*

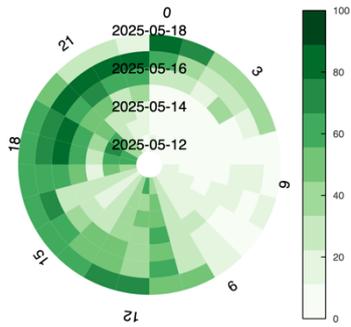 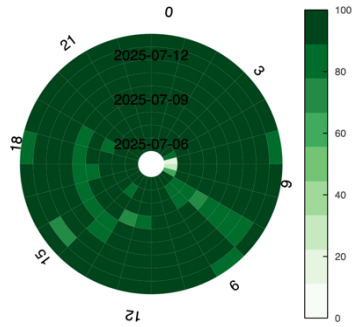 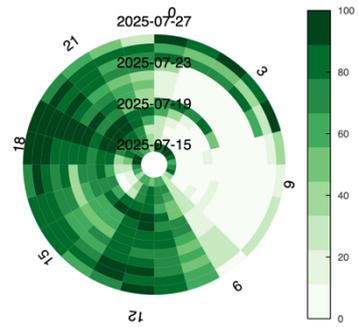

| *m) Human, May 12-18th* | *n) Human, July 6-14th* | *o) Human, 15-27th* |
|---|---|---|

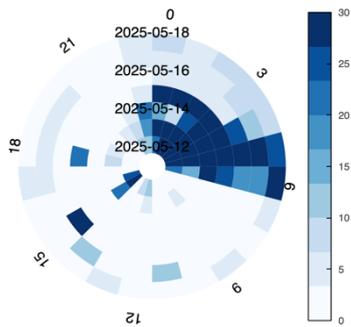 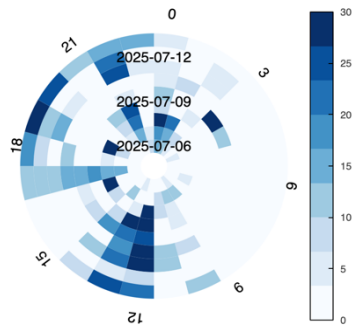 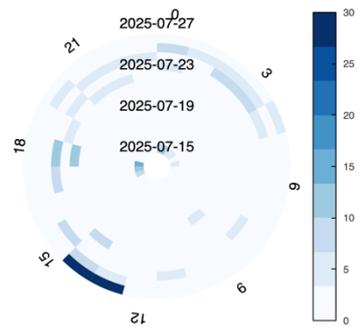

| *p) Music, May 12-18th* | *q) Music, July 6-14th* | *r) Music, July 15-27th* |
|---|---|---|

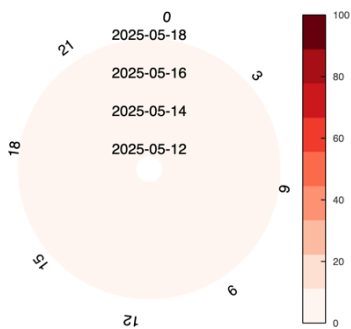 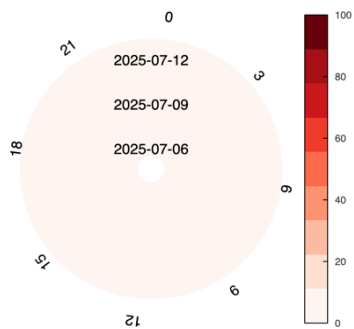 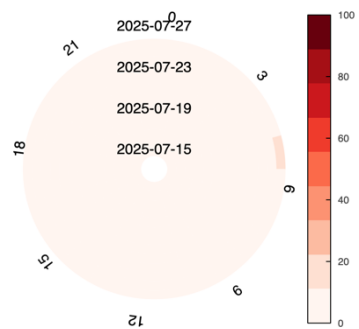

| *s) Vehicles, May 12-18th* | *t) Vehicles, July 6-14th* | *u) Vehicles, July 15-27th* |
|---|---|---|

Go to Results    Ir a Resultados

*Figure 16 - Sensor 05 circular plots across time and locations.*

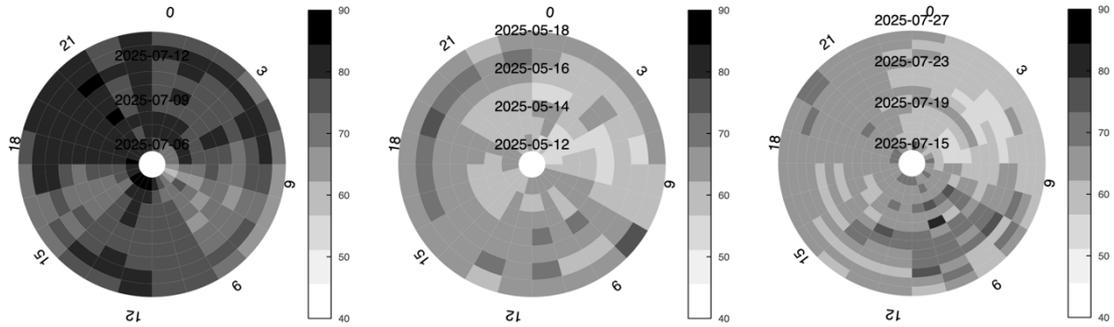

*a) LAeq, May 12-18th*    *b) LAeq, July 6-14th*    *c) LAeq, July 15-27th*

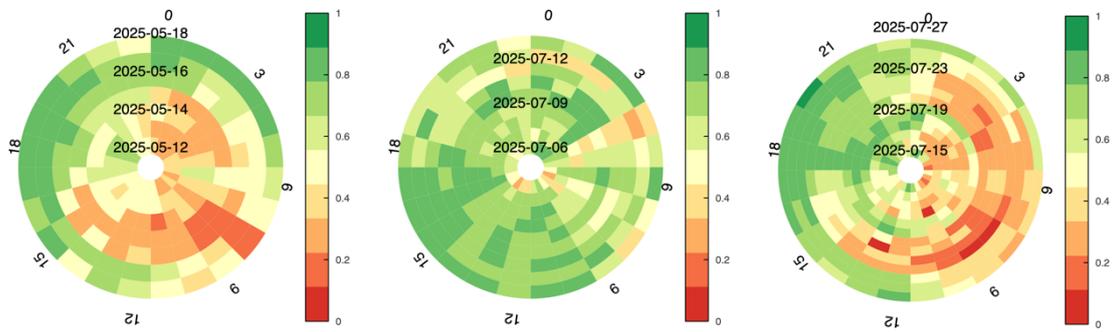

*d) Pleasantness, May 12-18th*    *e) Pleasantness, July 6-14th*    *f) Pleasantness, July 15-27th*

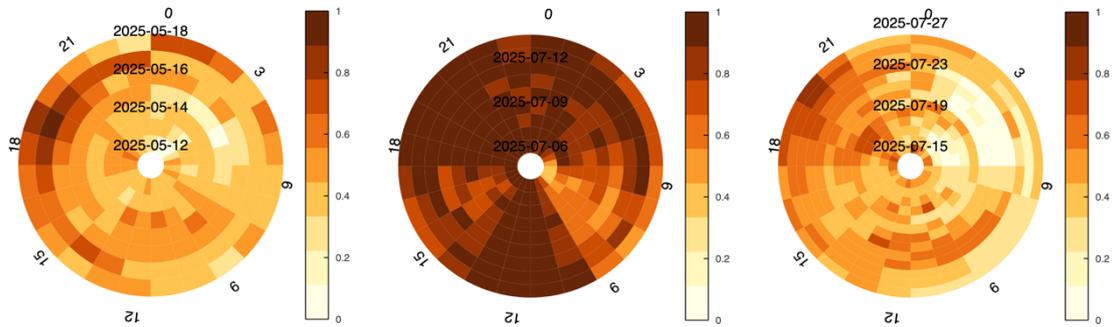

*g) Eventfulness, May 12-18th*    *h) Eventfulness, July 6-14th*    *i) Eventfulness, July 15-27th*

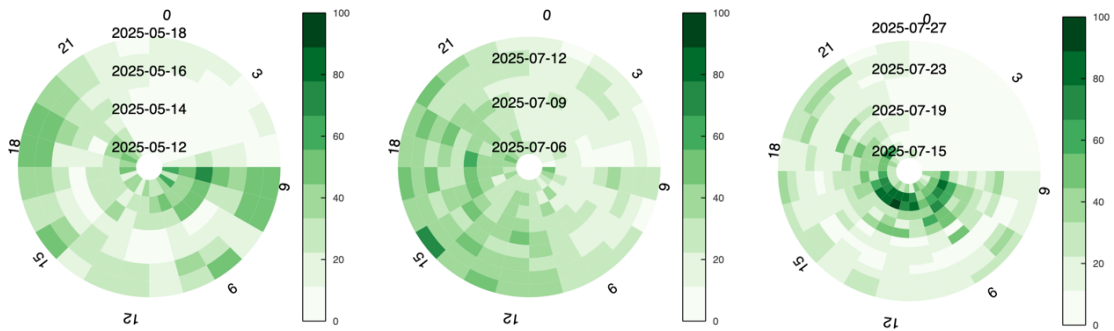

*j) Birds, May 12-18th*    *k) Birds, July 6-14th*    *l) Birds, July 15-27th*

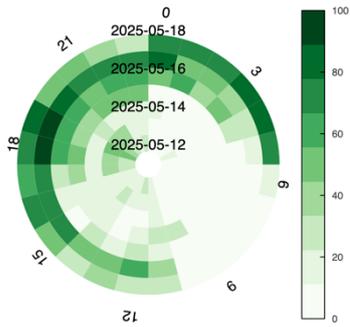
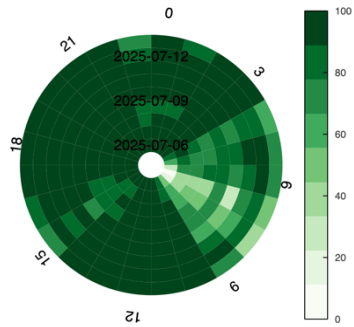
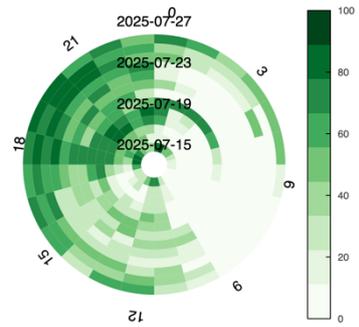

*m) Human, May 12-18th*     *n) Human, July 6-14th*     *o) Human, 15-27th*

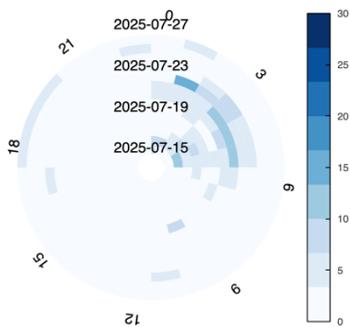
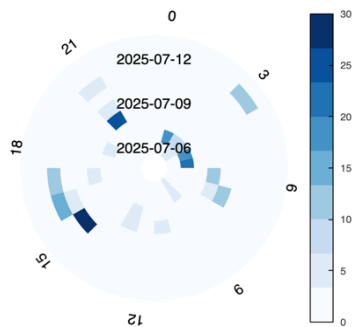
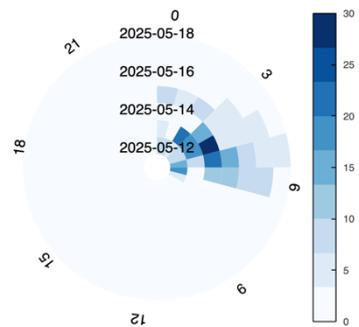

*p) Music, May 12-18th*     *q) Music, July 6-14th*     *r) Music, July 15-27th*

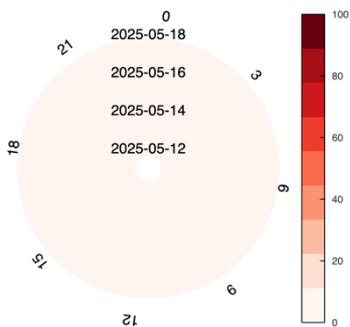
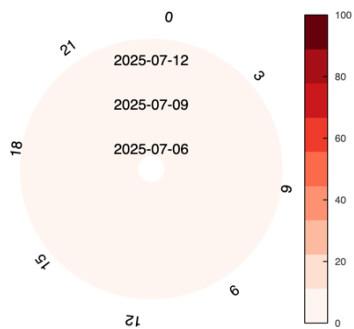
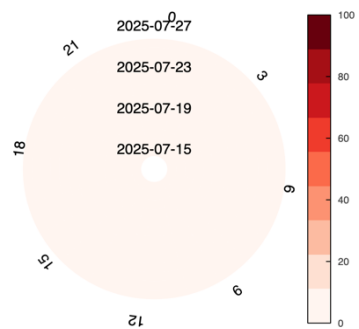

*s) Vehicles, May 12-18th*     *t) Vehicles, July 6-14th*     *u) Vehicles, July 15-27th*

Go to Results     Ir a Resultados